\documentclass{aastex631}

\usepackage{amsmath}
\usepackage{amsfonts}
\usepackage{diagbox}

\usepackage{chngcntr}
\shorttitle{Event-horizon-scale Imaging of M87* under Different Assumptions via Deep Generative Image Priors}
\shortauthors{Feng et al.}
\graphicspath{{./}{figures/}}

\begin{document}
\title{Event-horizon-scale Imaging of M87* under Different Assumptions via Deep Generative Image Priors}

\correspondingauthor{Berthy T. Feng}
\email{bfeng@caltech.edu}
\affiliation{California Institute of Technology (Caltech)}

\author{Berthy T. Feng}
\affiliation{California Institute of Technology (Caltech)}

\author{Katherine L. Bouman}
\affiliation{California Institute of Technology (Caltech)}

\author{William T. Freeman}
\affiliation{Massachusetts Institute of Technology (MIT)}
\affiliation{Google Research}

\begin{abstract}
Reconstructing images from the Event Horizon Telescope (EHT) observations of M87*, the supermassive black hole at the center of the galaxy M87, depends on a prior to impose desired image statistics. However, given the impossibility of directly observing black holes, there is no clear choice for a prior. We present a framework for flexibly designing a range of priors, each bringing different biases to the image reconstruction. These priors can be weak (e.g., impose only basic natural-image statistics) or strong (e.g., impose assumptions of black-hole structure). Our framework uses Bayesian inference with score-based priors, which are data-driven priors arising from a deep generative model that can learn complicated image distributions. Using our Bayesian imaging approach with sophisticated data-driven priors, we can assess how visual features and uncertainty of reconstructed images change depending on the prior. In addition to simulated data, we image the real EHT M87* data and discuss how recovered features are influenced by the choice of prior.
\end{abstract}

% \keywords{Astronomy image processing (2306) --- Supermassive black holes (1663)}


\section{Introduction}
\label{sec:intro}
% why imaging M87 is hard
In 2019, the Event Horizon Telescope (EHT) Collaboration obtained the first picture of M87* through computational imaging methods \citep{m87paperi,m87paperiv,m87papervi}. The published images gave humans a glimpse of the shadow cast by the supermassive black hole \citep{luminet1979image,Falcke_2000,Lu_2014} in the galaxy M87 based on data that EHT telescopes had collected in 2017 April \citep{m87paperii,m87paperiii}, and more recent images showed the persistence of the shadow one year later \citep{akiyama2024persistent}. However, these images necessarily incorporated imaging assumptions that were independent of telescope data. Because measurements obtained from very-long-baseline interferometry (VLBI) \citep{van1934wahrscheinliche,thompson2017interferometry} with EHT telescopes are corrupted and limited in number, infinitely many images --- many of them implausible and not interpretable --- would agree with a given set of measurements. Therefore, reconstructing an image from VLBI data requires assumptions about plausible image statistics in order to constrain the space of possible images \citep{m87paperiv}. 

% why we need to understand priors
Imaging assumptions can be formalized as a \textit{prior}, or a probability distribution of images that are acceptable regardless of observations \citep{scales2001prior}. More formally, we define a prior as a distribution of images $\mathbf{x}$ with a probability density function $p(\mathbf{x})$. In Bayesian inference the prior helps determine the image posterior $p(\mathbf{x}\mid\mathbf{y})$ given observations $\mathbf{y}$. However, designing a prior is not a straightforward task, especially considering the absence of true images of black holes. We address this problem with a principled strategy: we collect a range of priors that each impose different visual biases and plug these priors into a Bayesian imaging algorithm along with EHT VLBI data. Whereas the EHT Collaboration explored different imaging assumptions via the use of different imaging pipelines (e.g., CLEAN \citep{hogbom1974aperture,schwarz1978mathematical,clark1980efficient,schwab1984relaxing,cornwell1999deconvolution,shepherd2011difmap,Muller:2023xoo} and regularized maximum-likelihood (RML) methods \citep{bouman2016computational,chael2016high,akiyama2019smili}), we explore different priors within the same imaging pipeline. Our imaging approach allows us to assess how visual characteristics and uncertainty, as quantified through a Bayesian posterior, vary with the choice of prior.
An aim of our method is to easily move along the spectrum between strong constraints and weak constraints on the image. On one side of the spectrum lie model-fitting strategies, which find the parameters of an underlying geometric \citep{m87papervi,vincent2021geometric,nalewajko2020orientation,Lockhart_2022,sun2022alpha}, physical \citep{m87paperv,walsh2013m87,gebhardt2011black,Kawashima_2021,Yuan_2022,Nemmen_2019}, or statistical \citep{medeiros2023primo} model that best match the observations. On the other side lie traditional imaging approaches using weak regularizers like total variation \citep{akiyama2017imaging,kuramochi2018superresolution} and maximum entropy \citep{narayan1986maximum}. However, each side has its own limitations: model-fitting prevents discovering new features that cannot be explained by the assumed model, whereas traditional regularizers struggle to produce visually rich images. This motivates a method for imaging under a diverse array of priors, ranging from those akin to model-fitting (e.g., by assuming black-hole structure) to those similar to weak regularizers (e.g., by assuming basic properties of natural images). One can in principle obtain different priors by changing the regularization function \citep{muller2023using} or tuning regularization parameters \citep{m87paperiv}, but it is impractical to hand-design a regularizer for every desired prior.

A promising avenue is to use a data-driven prior that is fit to a training set of images with the desired statistics. Parameterizing the data-driven prior to be expressive enough is crucial. For example, principal components analysis (PCA) offers a simple probabilistic model of images, but it has been shown to not accurately express complicated image distributions \citep{feng2023score}. We choose to parameterize the prior by a score-based diffusion model (\textit{score-based prior}), which has been demonstrated as a powerful deep generative model capable of modeling a variety of image probability distributions \citep{song2021scorebased,ho2020denoising,dhariwal2021diffusion}, no matter how simple or complicated. That is, given a desired image prior $p(\mathbf{x})$, it can be assumed that the learned prior $p_\theta(\mathbf{x})$ with parameters $\theta$ trained on samples from $p(\mathbf{x})$ satisfies $p_\theta(\mathbf{x})\approx p(\mathbf{x})$ \citep{feng2023score}.

% our results
With score-based priors, we achieve a collection of M87* images that all fit the observed data but differ in certain visual characteristics. Specifically, we trained a score-based prior on each of the following datasets: CIFAR-10~\citep{cifar10} (generic natural images), general relativistic magneto-hydrodynamic (GRMHD) simulations~\citep{wong2022patoka}, radially inefficient accretion flow (RIAF) simulations~\citep{riaf}, and CelebA~\citep{celeba} (celebrity faces). We use a Bayesian imaging technique to apply each prior to the M87* observations, resulting in a collection of image posteriors. Each posterior is a probability distribution of images conditioned on the M87* data but incorporating a different prior. The visual biases of images from different posteriors are strikingly different, yet the images share structure that is prior-invariant, such as the ring shape and brightness asymmetry. We thus present two contributions based on our results: (1) a collection of possible M87* images that humans can selectively interpret based on their trust of the assumed biases and (2) analysis of which extracted black-hole features are robust to the prior and can be reliably used in scientific analysis.

% our findings
In this manuscript, we first describe relevant background in Section \ref{sec:background} and our Bayesian imaging method involving score-based priors in Section \ref{sec:method}. In Section \ref{sec:simdata_results}, we validate the imaging approach on simulated data using a collection of score-based priors ranging from weak biases (e.g., a prior trained on generic natural images) to strong biases (e.g., a prior trained on RIAF images). In Section \ref{sec:m87_results}, we present image posteriors of M87* based on the same collection of priors. Next, in Section \ref{sec:features}, we analyze the influence of the prior on characteristic ring features, including diameter, width, and orientation, by performing tests on both the simulated-data and M87* images. Finally, in Section \ref{sec:discussion}, we summarize our findings and conclude that our proposed imaging strategy allows one to define any collection of priors and analyze their effect on reconstructed images.

\section{Background}
\label{sec:background}

\subsection{Bayesian Imaging}
\label{subsec:bayesian_imaging}
Given measurements $\mathbf{y}=\mathbf{f}(\mathbf{x}^*)$, where $\mathbf{f}$ is a known forward model and $\mathbf{x}^*$ is an unknown source image, we would like to recover an image $\mathbf{x}$ such that $\mathbf{f}(\mathbf{x})\approx\mathbf{y}$. However, when $\mathbf{y}$ is sparse and corrupted, there is inherent uncertainty in the inverse problem of finding $\mathbf{x}$ \citep{scales2001prior}. Bayesian imaging accounts for this uncertainty by modeling a probability distribution known as the \textit{posterior}, or $p(\mathbf{x}\mid\mathbf{y})$. According to Bayes' rule, the log probability of an image under the posterior is given by
\begin{align}
\label{eq:log_posterior}
    \log p(\mathbf{x}\mid\mathbf{y})=\log p(\mathbf{y}\mid\mathbf{x})+\log p(\mathbf{x})+\text{const.}
\end{align}
We refer to $p(\mathbf{y}\mid\mathbf{x})$ as the measurement likelihood, or simply \textit{likelihood}, and we refer to $p(\mathbf{x})$ as the \textit{prior}. In this work, the posterior is based on a forward model for interferometric data and a score-based prior.

\subsection{Score-based Priors}
A score-based diffusion model is a deep generative model that learns to sample from an image distribution \citep{song2021scorebased,sohl2015deep,ho2020denoising}. We refer to its generative image distribution as a \textit{score-based prior} and denote it as $p_\theta$, where $\theta$ are the learned parameters. Given training images from a target prior $p_\text{data}$, the diffusion model is trained so that $p_\theta\approx p_\text{data}$.

Many methods have been developed to solve inverse problems with a pre-trained diffusion model \citep{song2022solving,chung2022come,chung2022score,choi2021ilvr,chung2022improving,chung2023diffusion,jalal2021robust,graikos2022diffusion,kawar2022denoising,adam2022posterior,song2023pseudoinverseguided,mardani2023variational,dia2023bayesian}. However, these methods typically produce a conditional distribution that does not correspond to an exact posterior. Since we instead prioritize posterior estimation, we turn to a more accurate approach that approximates the posterior through variational inference with a standalone score-based prior \citep{feng2023score,feng2023efficient}.

The crucial benefit of a score-based diffusion model is that it allows for computing image probabilities under $p_\theta$ \citep{song2021scorebased}. That is, given any image $\mathbf{x}$, we can compute $\log p_\theta(\mathbf{x})$ through an analytical, differentiable formula. To sample from a posterior whose prior is a score-based prior, we can use any posterior-sampling approach that requires the value or gradient of the posterior log density.

While $\log p_\theta(\cdot)$ has been used for posterior sampling~\citep{feng2023score}, the function is slow to compute and only feasible for images with at most $32\times 32$ pixels. We therefore appeal to a recently-proposed \textit{surrogate} score-based prior that is more computationally efficient~\citep{feng2023efficient}. The surrogate score-based prior is based on the evidence lower bound (ELBO) $b_\theta(\cdot)$ of a score-based prior. Instead of evaluating $\log p_\theta(\mathbf{x})$, we evaluate $b_\theta(\mathbf{x})\leq\log p_\theta(\mathbf{x})$ as the surrogate log density. Please refer to Appendix~\ref{app:score} for details about score-based diffusion models, including the formulae for $\log p_\theta(\cdot)$ and $b_\theta(\cdot)$.

\subsection{EHT Measurements}
The EHT performs VLBI with a global array of radio telescopes. Each pair of telescopes $i,j$, known as a \textit{baseline}, provides a Fourier measurement called a \textit{visibility} $v_{ij}$ \citep{van1934wahrscheinliche,zernike1938concept,thompson2017interferometry}. However, the baselines only sparsely sample the complex 2D Fourier plane, or $(u,v)$ space. Moreover, the visibilities are affected by thermal noise, station-dependent gain errors, and station-dependent phase errors \citep{m87paperiii}.

To overcome station-dependent errors, we use robust data products known as closure quantities. A \textit{closure phase} \citep{jennison1958phase} is given by a triplet of telescopes $i,j,k$ and computed as $\angle\left(v_{ij}v_{jk}v_{ki}\right)$. A \textit{log closure amplitude} \citep{twiss1960brightness} is given by a combination of four telescopes $i,j,k,\ell$ and computed as $\log\left(\frac{\lvert v_{ij}\rvert\lvert v_{k\ell}\rvert}{\lvert v_{ik}\rvert\lvert v_{j\ell}\rvert}\right)$. We denote the set of all linearly independent observed closure phases as $\mathbf{y}_\mathrm{cp}\in\mathbb{R}^{N_\mathrm{cp}}$ and that of log closure amplitudes as $\mathbf{y}_\mathrm{logca}\in\mathbb{R}^{N_\mathrm{logca}}$. In the case of visibilities with a high signal-to-noise ratio (i.e., SNR $>1$), closure phases and log closure amplitudes approximately experience mean-zero Gaussian thermal noise \citep{rogers1995fringe,Blackburn_2020,Broderick_2020} with standard deviations $\mathbf{\sigma}_\mathrm{cp}\in\mathbb{R}^{N_\mathrm{cp}}$ and $\mathbf{\sigma}_\mathrm{logca}\in\mathbb{R}^{N_\mathrm{logca}}$, respectively. We assume the high-SNR setting. Conditioned on an image $\mathbf{x}$, the measurement distribution can be modeled as Gaussian with log likelihoods
\begin{align}
\label{eq:llh}
    \log p(\mathbf{y}_\text{cp}\mid\mathbf{x}) = -\frac{1}{2\mathbf{\sigma}_\text{cp}^2}\left\lVert\mathbf{f}_\text{cp}(\mathbf{x})-\mathbf{y}_\text{cp}\right\rVert_2^2 \quad\text{and}\quad \log p(\mathbf{y}_\text{logca}\mid\mathbf{x}) = -\frac{1}{2\mathbf{\sigma}_\text{logca}^2}\left\lVert\mathbf{f}_\text{logca}(\mathbf{x})-\mathbf{y}_\text{logca}\right\rVert_2^2,
\end{align}
where $\mathbf{f}_\mathrm{cp}$ and $\mathbf{f}_\mathrm{logca}$ are nonlinear forward models. We note that the Gaussian noise is not purely (statistically) independent, but it is approximately independent under high-SNR visibilities or can be made independent under a linear transformation~\citep{Blackburn_2020,Broderick_2020,arras2022variable}. Please refer to Appendix \ref{app:interferometry} for details about interferometric data products and their forward models.

It is possible to use the same imaging algorithm with visibility amplitudes instead of log closure amplitudes. Visibility amplitudes, which have been used for other imaging results~\citep{m87paperiv,medeiros2023primo}, are more constraining than closure amplitudes, but they require calibration according to assumptions such as station-dependent systematic noise. In this work, we focus on using log closure amplitudes in order to avoid tuning the calibration assumptions. The original M87* work includes reconstructions from both types of data products for reference~\citep{m87paperiv}.

\newpage
\section{Method}
\label{sec:method}
\begin{figure}[t]
    \centering
    \includegraphics[width=0.8\textwidth]{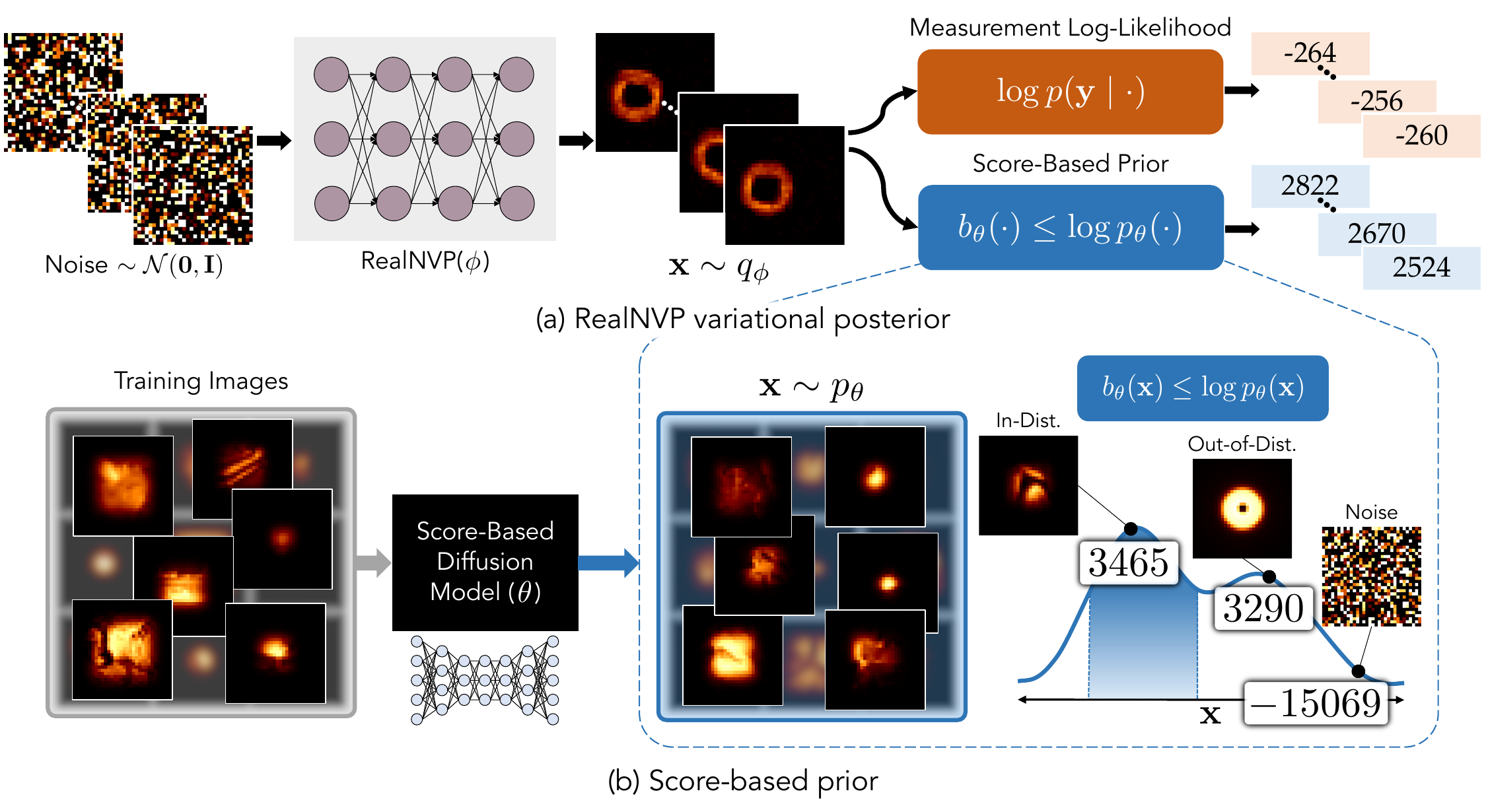}
    \caption{Method illustration. The CIFAR-10 prior was used for these examples; images are shown as $32\times 32$ pixels on a $[0,1]$ scale. At a high level, we optimize a variational distribution $q_\phi$ to approximate the image posterior $p_\theta(\cdot\mid\mathbf{y})$ given a score-based prior $p_\theta$ and log likelihood based on EHT measurements. Panel (a) illustrates our particular variational distribution: a RealNVP with parameters $\phi$. At each optimization iteration $i$, the measurement log likelihood (Equation \eqref{eq:llh}) and the log density under the score-based prior of each sample $\mathbf{x}$ from $q_\phi= q_{\phi^{(i)}}$ are evaluated. The average gradient is computed with respect to $\phi$ to update $\phi^{(i)}$. In other words, $q_\phi$ is optimized to produce samples that have high probability under both measurement likelihood and prior. Panel (b) zooms in to the score-based prior. A score-based prior is based on a score-based diffusion model, a deep generative model with parameters $\theta$, that is trained on images from a target prior. Once trained, the diffusion model generates samples from a generative image distribution $p_\theta$. There is an analytical formula for computing the ELBO $b_\theta(\mathbf{x})$ of the log probability $\log p_\theta(\mathbf{x})$ for any image $\mathbf{x}$, even for out-of-distribution images and images of pure noise.}
    \label{fig:method}
\end{figure}

Here we describe how to sample from an image posterior given EHT measurements and a score-based prior. The method is that of \citet{feng2023efficient} with a measurement likelihood based on closure quantities (Equation \eqref{eq:llh}). We formulate the following posterior log density:
\begin{align}
\label{eq:posterior_objective}
    \log p_\theta(\mathbf{x}\mid\mathbf{y})&=\log p(\mathbf{y}_\text{cp}\mid\mathbf{x})+\log p(\mathbf{y}_\text{logca}\mid\mathbf{x})+\log p_\theta(\mathbf{x})- \left(V(\mathbf{x})-\bar{V}\right)^2+\text{const.},
\end{align}
where $p_\theta(\mathbf{x}\mid\mathbf{y})$ is given a $\theta$ subscript to clarify its dependence on the score-based prior $p_\theta$.
We include a flux-constraint objective $-\left(V(\mathbf{x})-\bar{V}\right)^2$, where $V(\mathbf{x})$ is the total flux (i.e., the sum of the pixel values) of the image $\mathbf{x}$ and $\bar{V}$ is the target total flux, because closure quantities do not constrain the total flux. We set $\bar{V}$ as the median total flux of images sampled from the score-based prior $p_\theta$ and then scale posterior images to the original total flux as measured in the zero-baseline visibility. Please see Appendix~\ref{app:ablation} for a discussion on the flux-constraint objective. Note that since our priors were trained on compact centered images, we do not need an explicit center-of-light constraint.

We use variational inference to sample from $p_\theta(\mathbf{x}\mid\mathbf{y})$, approximating the complicated posterior with a tractable distribution known as the variational distribution. We use a RealNVP \citep{dinh2016density}, a type of deep generative model known as a \textit{normalizing flow} \citep{papamakarios2021normalizing}, with parameters $\phi$ as the variational distribution \citep{sun2021deep}. Samples from the RealNVP are images $\mathbf{x}$ from a distribution $q_\phi$, and we optimize $\phi$ to minimize an upper bound on the Kullback-Leibler (KL) divergence $D_\text{KL}(q_\phi\lVert p_\theta(\cdot\mid\mathbf{y}))$:
\begin{align}
\label{eq:objective}
    \phi^*=\arg\min_\phi \mathbb{E}_{\mathbf{x}\sim q_\phi} \left[-\log p(\mathbf{y}_\text{cp}\mid\mathbf{x})-\log p(\mathbf{y}_\text{logca}\mid\mathbf{x})-b_\theta(\mathbf{x})+\left(V(\mathbf{x})-\bar{V}\right)^2+\log q_\phi(\mathbf{x})\right],
\end{align}
where $b_\theta(\mathbf{x})\leq\log p_\theta(\mathbf{x})$ is an efficient lower bound on the exact log probability of $\mathbf{x}$ under the score-based prior~\citep{feng2023efficient,song2021maximum}.
% The term $\log q_\phi(\mathbf{x})$ mathematically arises from the KL objective and can be interpreted as an entropy objective preventing mode collapse.

We approximately solve Equation \eqref{eq:objective} with stochastic gradient descent, iteratively Monte-Carlo approximating the KL upper bound with a batch of samples from $q_\phi$ and computing the gradient with respect to $\phi$.

We find that annealing the weight of the data-fit terms gradually from $0$ to $1$ helps prevent bad local minima (see Appendix \ref{app:data_annealing} for a discussion on data annealing). Furthermore, we find that optimization can be sensitive to the chosen target flux $\bar{V}$ and data annealing schedule. One way to mitigate this is to make sure the diffusion model has a median total flux that is close to the median total flux of the training images. The data annealing may need to be tuned to achieve a local minimum at which the posterior images exhibit characteristic features of the prior (e.g., posterior images should be centered if all the training images for the prior are centered).
Once $\phi$ is optimized, samples $\mathbf{x}\sim q_\phi$ can be efficiently obtained as samples from an approximate posterior. The RealNVP occasionally outputs slightly negative pixel values, so we clip samples at inference time to a minimum value of $0$ to impose a positivity constraint. Please refer to Appendix~\ref{app:implementation} for implementation details. Figure \ref{fig:method} illustrates the essential components of our imaging method: the RealNVP variational posterior and the score-based prior.

\subsection{Score-based Priors Used in This Work}
\label{subsec:priors}
\begin{figure}[ht]
    \centering
    \includegraphics[width=0.7\textwidth]{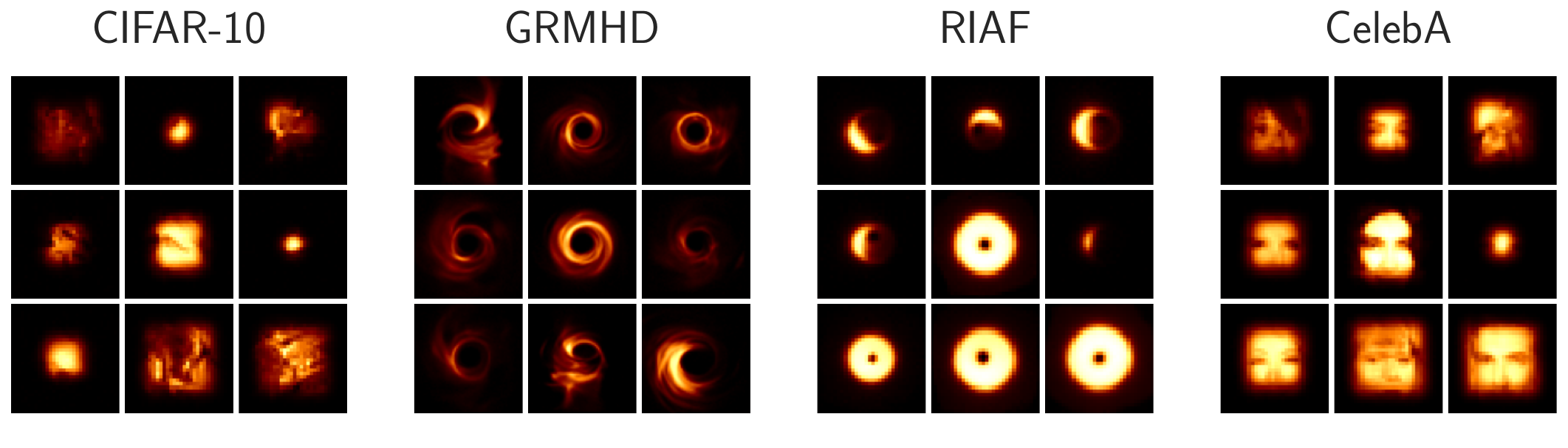}
    \caption{Score-based priors used in this work. Nine samples from each learned prior are shown.}
    \label{fig:priors}
\end{figure}

In this work, we focus on the following score-based priors, each trained on a dataset of images assuming a $128$-$\mu$as field of view (FOV). Unless stated otherwise, we use the terms ``prior'' and ``score-based prior'' synonymously. 
Figure \ref{fig:priors} shows samples from each prior.
\begin{itemize}
    \item The \textbf{CIFAR-10} prior was trained on the CIFAR-10~\citep{cifar10} dataset of $32\times 32$ images from $10$ object classes (e.g., airplane, automobile, dog). We used a training set of $45$K grayscale images. The images were tapered on the edges to incorporate assumptions of a black background and a centered object. A tapering effect is created by defining a binary mask with a center square region of pixels set to $1$ and everywhere else set to $0$, then applying a Gaussian blur kernel with standard deviation $8$ $\mu$as, and then element-wise multiplying the blurred mask with the image. The size of the taper was randomly varied during training by randomly varying the size of the center square region of the mask, resulting in a centered compact region of between $12.8\times 12.8$ and $83.2\times 83.2$ $\mu$as.
    \item The \textbf{GRMHD} prior was trained on $100$K images from GRMHD simulations~\citep{wong2022patoka} of Sgr A* resized to $64\times 64$ pixels. During training, the GRMHD images were randomly flipped horizontally and randomly zoomed between $-16.7\%$ (zoomed-in) and $+14.5\%$ (zoomed-out), thus varying the diameter of the thin ring to be between $35$ and $48$ $\mu$as.
    \item The \textbf{RIAF} prior was trained on $9070$ images of RIAF \citep{riaf} simulations. The RIAF images were downloaded\footnote{\url{http://vlbiimaging.csail.mit.edu/myData}} with all available spin and inclination parameters and resized to $32\times 32$. During training, they were also randomly zoomed between $-16.7\%$ and $+14.5\%$, randomly flipped horizontally, and randomly rotated between $-2\pi$ and $+2\pi$.
    \item The \textbf{CelebA} prior was trained on the CelebA~\citep{celeba} dataset of celebrity faces. We used a training set of $160$K images that were resized to $32\times 32$. The same tapering effect that was used for CIFAR-10 was applied. Although far from a reasonable prior for astronomical images, a prior trained on faces helps us see what happens when strong but probably incorrect assumptions are made.
\end{itemize}
The RIAF and GRMHD priors incorporate strong assumptions about a ring structure. The CIFAR-10 and CelebA priors, on the other hand, do not assume any ring structure or even the presence of an astronomical object. One might make the following order of priors from weak to strong assumptions: CIFAR-10, GRMHD, RIAF. In addition, we have the CelebA prior, which makes specific assumptions against our expectations.

\section{Simulated-data Results}
\label{sec:simdata_results}
We validate our imaging approach on simulated observations of synthetic source images. A crucial advantage of our approach is that it does not involve parameter-tuning. As we do not need to hand-tune hyperparameters based on a calibration dataset, the experiments presented in this section are simply meant to verify the efficacy of the approach.

\begin{figure}[ht]
    \centering
    \includegraphics[width=0.7\textwidth]{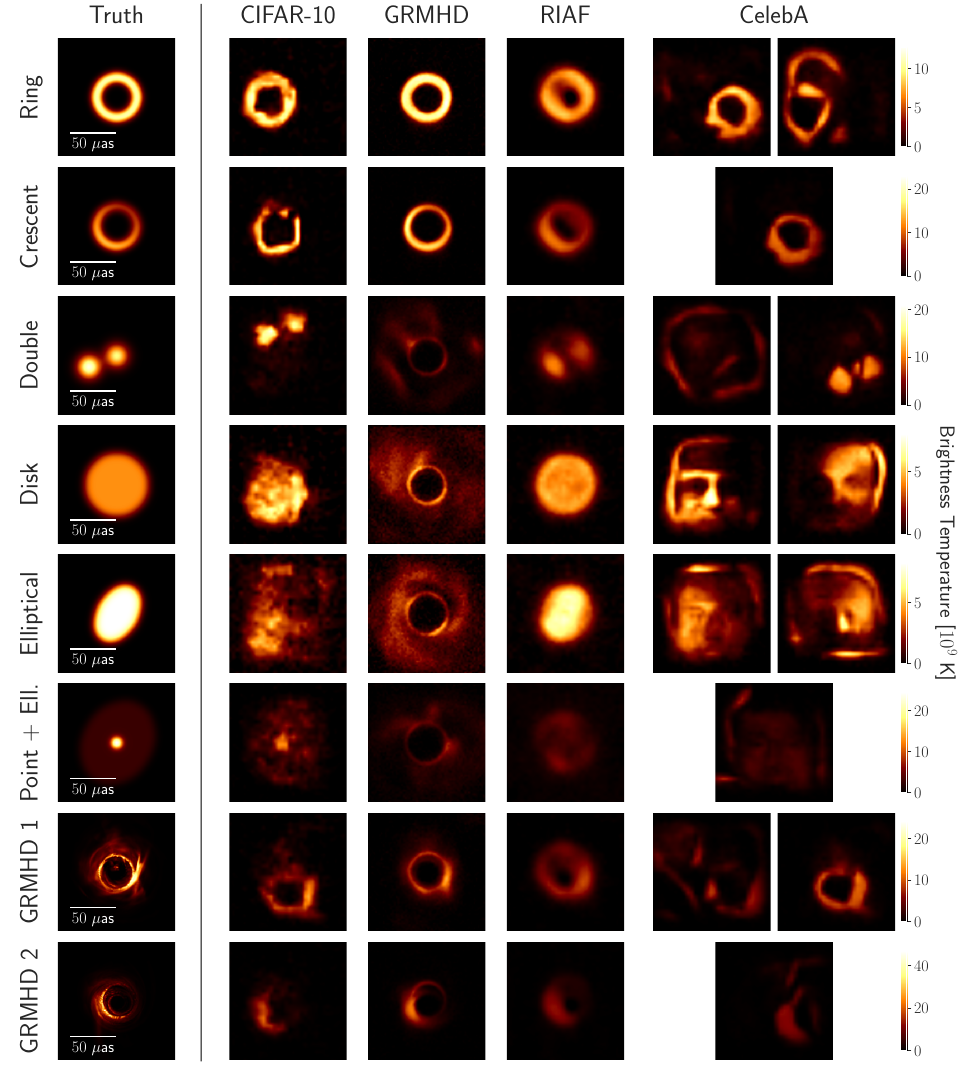}
    \caption{Image reconstructions from simulated data. A sample (one sample from each mode if the posterior is bimodal) is shown from each estimated posterior. Qualitatively, the CIFAR-10 prior adds the least amount of bias, producing reasonable reconstructions of each image in this dataset. The GRMHD prior strongly prefers a centered ring in the image. The RIAF prior prefers a centered ring- or disk-like structure in the image. The CelebA prior struggles to recover these source images, in some cases adding face features, and it leads to the most multimodal posteriors. However, it performs decently well on certain images like the Crescent and GRMHD images. When the source image is known to be well-approximated by a GRMHD or RIAF model, the more constrained GRMHD or RIAF prior may be the best choice.}
    \label{fig:simdata}
\end{figure}

Figure \ref{fig:simdata} shows results for a dataset of eight source images. Among the images are validation images used in the original M87* imaging work~\citep{m87paperiv} and two images of an elliptical object used in the Sgr A* imaging work~\citep{sgrapaperiii}. All observations were simulated based on the April 6 observing array using code provided by \citet{m87paperiv}. We used closure phases and log closure amplitudes of the combined low-band and high-band data and followed the same preprocessing steps as the \texttt{eht-imaging} algorithm (assuming non-closing fractional systematic noise of $0.03$), except we did not add station-dependent systematic noise since we do not need to calibrate the visibility amplitudes. Although imaging was done with a prior-dependent total flux and either $32\times 32$ or $64\times 64$ pixels depending on the prior, we re-scale images to have a total flux of $0.6$ Jy and resize them to $128\times 128$ for visualization.

The quality of image reconstruction heavily depends on the prior. For example, the GRMHD reconstruction of GRMHD 1 appears more convincing than the GRMHD reconstruction of the Double image in Figure \ref{fig:simdata}. On the other hand, the RML methods used in previous EHT imaging efforts \citep{m87paperiv,sgrapaperiii} achieve overall cleaner reconstructions of synthetic data. One reason for the better performance of those RML methods is that they use regularization parameters chosen based on a calibration dataset that is very similar to their test images. In contrast, we consider priors that may be profoundly different from the true source image (e.g., CelebA prior applied to the Double data). Another reason is that RML methods produce a mean image that tends to be cleaner than individual posterior samples, which are shown in Figure \ref{fig:simdata}. We emphasize that the goal of our work is not to achieve the cleanest or most accurate reconstructions; rather, we aim to showcase the effects of different priors, even when those priors might not lead to the best reconstruction due to mismatch with the data.

Overall, the reconstructed images make sense according to the biases of the prior. The CelebA prior introduces face-like features, especially when the ground-truth source object is fairly ``flat'' (e.g., the Disk and Elliptical images), and it is the only prior that leads to multimodal estimated posteriors. Images under the RIAF prior are always centered and ring- or disk-like. The GRMHD prior always prefers the presence of a thin ring at the center of the image. The CIFAR-10 prior imposes weak biases and appears to assemble images from small, locally-smooth patches.
% The CIFAR-10 prior imposes weak image biases and appears to assemble images from small image patches, which can be understood by considering that natural images have been shown to be well-approximated by patch-based priors~\citep{zoran2012natural}. 

Table \ref{tab:simdata_ncc} quantifies the performance of the various priors on each source image. As in previous work \citep{m87paperiv,sgrapaperiii}, we evaluate the normalized cross-correlation (NCC). Since our approach does not explicitly constrain the center-of-light, we use a shift-invariant NCC metric, which is computed as the maximum NCC between all shifted versions of the reconstructed image and the ground-truth image. As expected, the closer the prior is to the ground-truth image, the more accurately it recovers the ground-truth image. For example, the GRMHD prior excels at recovering the Ring, Crescent, and GRMHD images but struggles with the non-ring-like images. The RIAF prior performs well on ring-like images and disk-like images. The CelebA prior, unsurprisingly, performs poorly on this dataset of images. The CIFAR-10 prior does best compared to the other priors at recovering the non-ring-like images (e.g., Double and Point+Elliptical). It performs generally well across all the source images, suggesting that it serves as an effective ``general-purpose'' prior. 

Table \ref{tab:simdata_chisq} quantifies agreement with the simulated data using the reduced $\chi^2$ metric --- we note that it is not a true reduced $\chi^2$ since we only incorporate image pixels as degrees of freedom, but it is useful as a proxy metric of data consistency. There does not appear to be a correlation between the $\chi^2$ statistics in Table \ref{tab:simdata_chisq} and the NCC statistics in Table \ref{tab:simdata_ncc}. The results in Table \ref{tab:simdata_chisq} simply confirm data consistency of the reconstructed images, as $\chi^2$ values are consistently less than $2$ and often close to $1$ (lower $\chi^2$ corresponds to more data consistency, and $\chi^2\approx 1$ is considered a sign of a good balance between data and prior). The RIAF prior results in the highest $\chi^2$ values, perhaps because it is the most constraining prior. Overall, our tests on simulated data confirm that the score-based priors impose the expected biases on the image reconstruction while allowing for reasonable data consistency.

\begin{deluxetable*}{rCCCC}
\label{tab:simdata_ncc}
\tablewidth{0pt}
\tablehead{
\nocolhead{Image} & \colhead{CIFAR-10} & \colhead{GRMHD} & \colhead{RIAF} &
\colhead{CelebA}
}
\startdata
Ring & 0.95 \pm 0.009 & \mathbf{0.98} \pm 0.024 & 0.90 \pm 0.003 & 0.88 \pm 0.090 \\ 
Crescent & 0.85 \pm 0.019 & \mathbf{0.96} \pm 0.007 & 0.92 \pm 0.005 & 0.95 \pm 0.007 \\ 
Double & \mathbf{0.94} \pm 0.009 & 0.49 \pm 0.028 & 0.93 \pm 0.004 & 0.39 \pm 0.130 \\ 
Disk & 0.95 \pm 0.007 & 0.59 \pm 0.027 & \mathbf{0.99} \pm 0.002 & 0.80 \pm 0.020 \\ 
Elliptical & 0.79 \pm 0.037 & 0.50 \pm 0.029 & \mathbf{0.97} \pm 0.002 & 0.75 \pm 0.026 \\ 
Point + Ell. & \mathbf{0.87} \pm 0.022 & 0.58 \pm 0.031 & 0.68 \pm 0.010 & 0.43 \pm 0.026 \\ 
GRMHD 1 & 0.84 \pm 0.015 & \mathbf{0.86} \pm 0.009 & 0.83 \pm 0.004 & 0.46 \pm 0.089 \\ 
GRMHD 2 & 0.85 \pm 0.012 & \mathbf{0.90} \pm 0.003 & 0.85 \pm 0.006 & 0.77 \pm 0.012 \\ 
\enddata
\caption{Normalized cross-correlation (NCC). The avg.~$\pm$ std.~dev.~of the NCC metric for $128$ samples from the posterior is reported (highest NCC in each row is shown in bold). In general, the closer the prior is to the ground-truth image, the closer its posterior samples are to the ground truth. For example, the Ring, Crescent, and GRMHD images are best reconstructed with the GRMHD prior, whereas non-ring-like images are poorly reconstructed with the GRMHD prior. The CIFAR-10 prior may be the best ``general-purpose'' prior, giving NCC values between about $0.80$ and $0.95$ across these images.}
\end{deluxetable*}

\begin{deluxetable*}{rLCCCCC}
\label{tab:simdata_chisq}
\tablewidth{0pt}
\tablehead{
\nocolhead{Image} & \nocolhead{Data Product} & \colhead{CIFAR-10} & \colhead{GRMHD} & \colhead{RIAF} & \colhead{CelebA}
}
\startdata
Ring & \chi^2_\mathrm{cp} & 0.87 \pm 0.02 & 0.90 \pm 0.03 & 0.96 \pm 0.02 & 0.89 \pm 0.04 \\ 
& \chi^2_\mathrm{logca} & 0.72 \pm 0.03 & 0.75 \pm 0.05 & 1.04 \pm 0.06 & 0.72 \pm 0.03 \\ \hline
Crescent & \chi^2_\mathrm{cp} & 0.73 \pm 0.02 & 0.76 \pm 0.02 & 0.84 \pm 0.03 & 0.73 \pm 0.02 \\ 
& \chi^2_\mathrm{logca} & 0.67 \pm 0.02 & 0.78 \pm 0.03 & 0.95 \pm 0.05 & 0.68 \pm 0.02 \\ \hline
Double & \chi^2_\mathrm{cp} & 0.96 \pm 0.02 & 1.00 \pm 0.03 & 1.01 \pm 0.03 & 0.99 \pm 0.03 \\ 
& \chi^2_\mathrm{logca} & 0.77 \pm 0.02 & 0.81 \pm 0.04 & 1.31 \pm 0.08 & 0.84 \pm 0.03 \\ \hline
Disk & \chi^2_\mathrm{cp} & 1.82 \pm 0.04 & 1.83 \pm 0.05 & 1.80 \pm 0.03 & 1.81 \pm 0.05 \\ 
& \chi^2_\mathrm{logca} & 1.30 \pm 0.04 & 1.56 \pm 0.14 & 1.41 \pm 0.04 & 1.42 \pm 0.06 \\ \hline
Elliptical & \chi^2_\mathrm{cp} & 1.76 \pm 0.04 & 1.76 \pm 0.04 & 1.91 \pm 0.04 & 1.79 \pm 0.04 \\ 
& \chi^2_\mathrm{logca} & 1.44 \pm 0.03 & 1.64 \pm 0.06 & 1.64 \pm 0.05 & 1.43 \pm 0.03 \\ \hline
Point + Ell. & \chi^2_\mathrm{cp} & 1.22 \pm 0.02 & 1.21 \pm 0.02 & 1.42 \pm 0.04 & 1.24 \pm 0.02 \\ 
& \chi^2_\mathrm{logca} & 0.83 \pm 0.02 & 0.83 \pm 0.03 & 0.96 \pm 0.03 & 0.84 \pm 0.02 \\ \hline
GRMHD 1 & \chi^2_\mathrm{cp} & 0.90 \pm 0.02 & 0.89 \pm 0.07 & 1.02 \pm 0.04 & 0.90 \pm 0.05 \\ 
& \chi^2_\mathrm{logca} & 0.72 \pm 0.03 & 0.70 \pm 0.03 & 1.08 \pm 0.07 & 0.83 \pm 0.03 \\ \hline
GRMHD 2 & \chi^2_\mathrm{cp} & 0.59 \pm 0.02 & 0.59 \pm 0.01 & 0.62 \pm 0.04 & 0.60 \pm 0.02 \\ 
& \chi^2_\mathrm{logca} & 0.51 \pm 0.02 & 0.56 \pm 0.03 & 1.01 \pm 0.08 & 0.53 \pm 0.02 
\enddata
\caption{Data-consistency metrics ($\chi^2$) for closure quantities of simulated data. $\chi_\mathrm{cp}$ and $\chi_\mathrm{logca}$ are the $\chi^2$ metrics for closure phases and log closure amplitudes, respectively. The avg.~$\pm$ std.~dev.~of $128$ samples from the estimated posterior is reported. Lower $\chi^2$ is a sign of higher data consistency. $\chi^2\approx 1$ is considered an indication of a good balance between the observed data and the prior. The Disk, Elliptical, and Point+Elliptical images are the most challenging cases for these particular priors, as evidenced by the high $\chi^2$ values that indicate data-fitting difficulty.}
\end{deluxetable*}

\subsection{Characterizing the Simulated-data Posteriors}
\begin{figure}[ht]
    \centering
    \includegraphics[width=\textwidth]{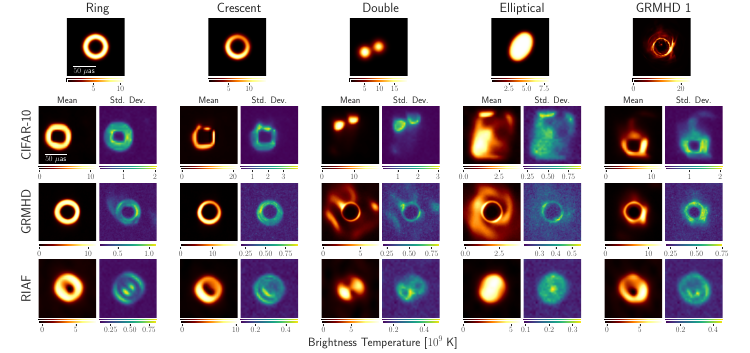}
    \caption{Mean and std.~dev.~of simulated-data posteriors. We find that uncertainty decreases with a stronger prior (i.e., maximum std.~dev.~decreases from CIFAR-10 to GRMHD to RIAF). CIFAR-10 exhibits the most uncertainty given that it is the most flexible of the priors. Compared to the individual samples in Figure \ref{fig:simdata}, the mean images appear much cleaner.}
    \label{fig:simdata_posterior}
\end{figure}

In addition to evaluating single samples from the posterior (Figure \ref{fig:simdata}), we can assess aspects of the posterior distribution such as uncertainty and multimodality. Figure \ref{fig:simdata_posterior} shows the mean and pixel-wise standard deviation of posteriors under the CIFAR-10, GRMHD, and RIAF priors. We find that uncertainty decreases as the prior becomes stronger. For a weak prior like CIFAR-10, which leads to high posterior uncertainty, it can be helpful to consider the mean reconstruction instead of noisier individual samples.

\begin{figure}[ht]
    \centering
    \includegraphics[width=0.85\textwidth]{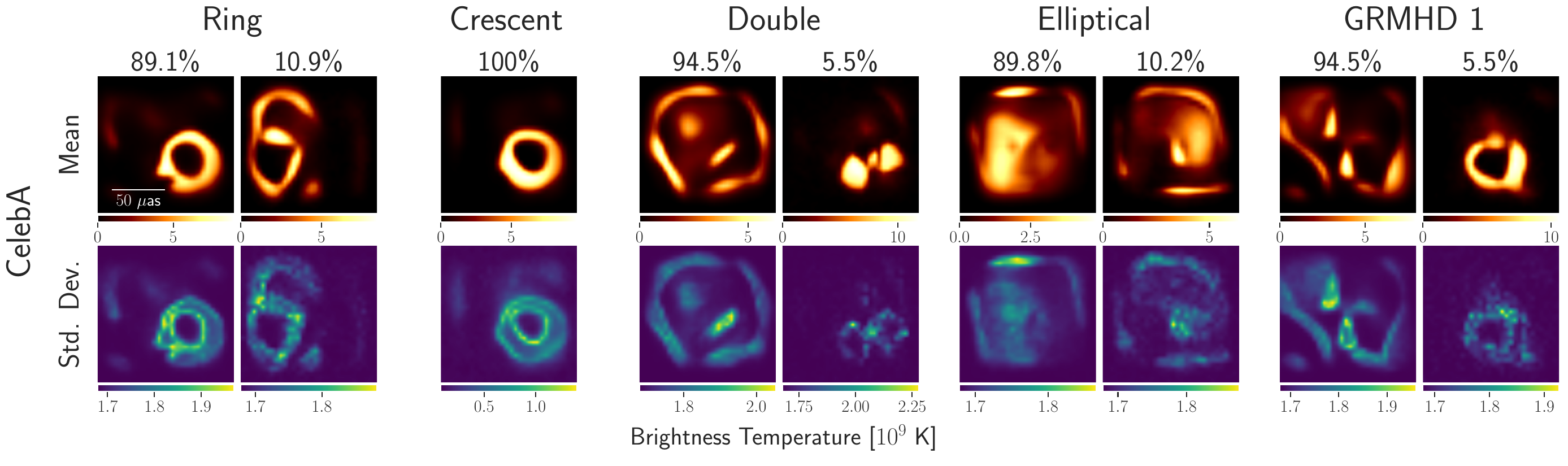}
    \caption{Mean and std.~dev.~of simulated-data posteriors under the CelebA prior. The CelebA prior is the only one that gives rise to bimodal posteriors. For each posterior, a one- or two-component Gaussian mixture model (one for a single-mode posterior and two for a bimodal posterior) was fit to $128$ samples. The mean, pixel-wise std.~dev., and weight of each Gaussian component are shown. For the Ring, Double, and GRMHD 1 data, one of the two modes is quite similar to the ground-truth image, while the other is not.}
    \label{fig:simdata_celeb}
\end{figure}

The CelebA prior leads to bimodal posteriors, which are characterized in Figure \ref{fig:simdata_celeb}. We note that the number of modes in the estimated posterior may be due to the variational family being used; in the future, a more expressive family of distributions parameterized through a better network architecture may identify more modes. It is perhaps reassuring that a prior with such erroneous assumptions (i.e., that face statistics well describe these particular source images) is able to account for mismatches with the data through a wider posterior. For example, for the Ring, Double, and GRMHD 1 data, the posterior covers two modes, one of which accurately recovers the ground-truth image.

\subsection{Biases of the CIFAR-10 Prior}
\begin{figure*}
\gridline{\fig{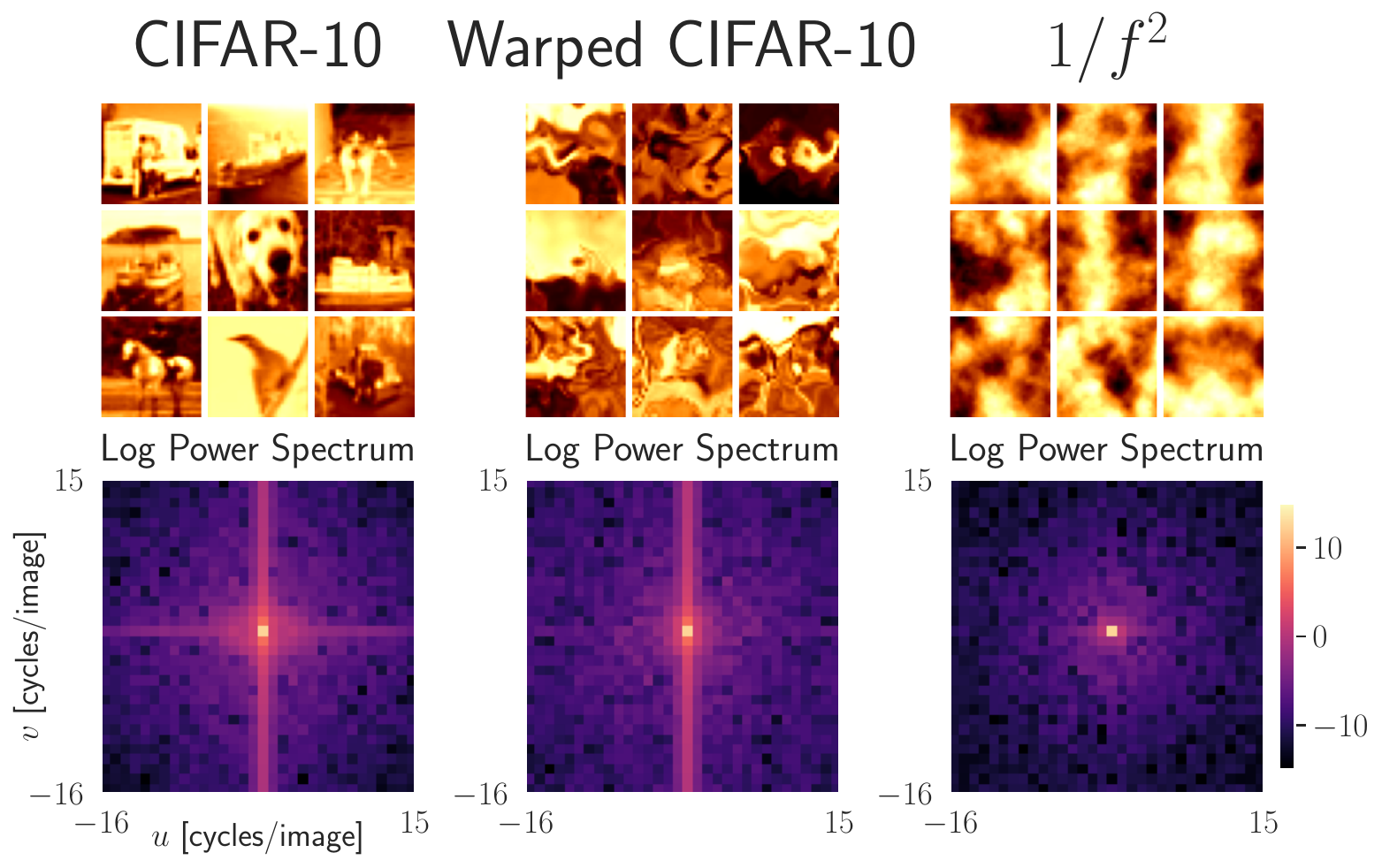}{0.38\linewidth}{(a) Image statistics of CIFAR-10 and alternative natural-image datasets}}\label{subfig:boxyness_statistics}
\gridline{\fig{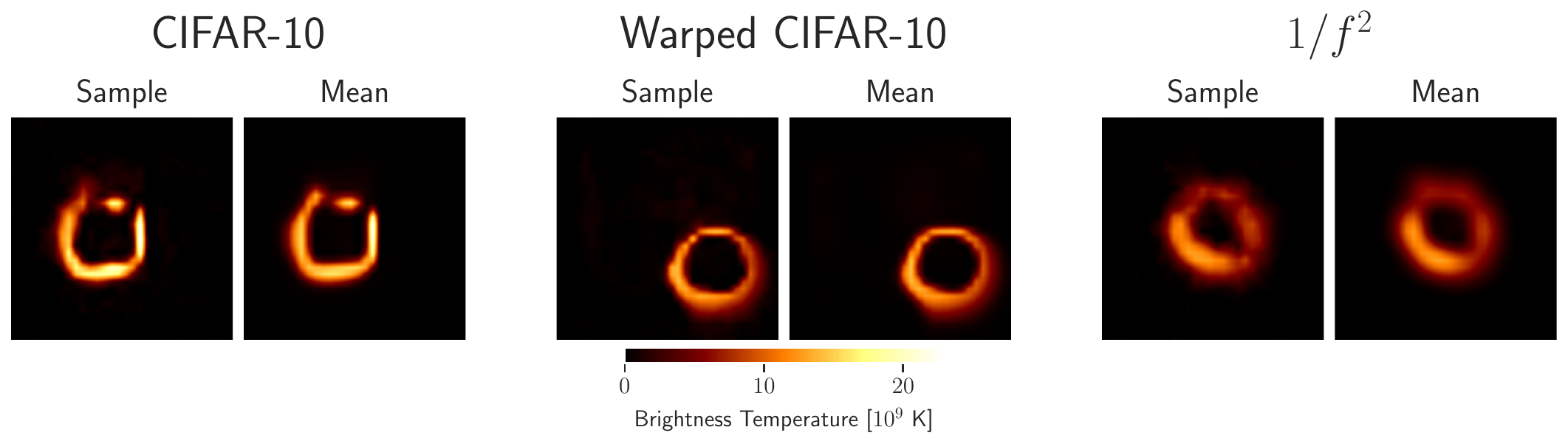}{0.6\linewidth}{(b) Crescent reconstructions under alternative natural-image priors}}
\caption{Reducing the CIFAR-10 bias toward horizontal lines and vertical lines. For example, given simulated data of the Crescent image from Figure \ref{fig:simdata}, this bias results in posterior images containing a boxy object, even though the actual Crescent is a ring without any straight edges. This is likely due to the statistics of the CIFAR-10 dataset, which contains mostly man-made objects and animals. Such images include many sharp corners and lines arising from boxy objects like cars, the legs of standing animals, or horizon lines. Although CIFAR-10 is our choice of a generic natural-image prior, it is possible to define an alternative natural-image prior with less preference for vertically- or horizontally-oriented edges. We tested a prior trained on CIFAR-10 images that underwent warped random affine transforms and a prior trained on $32\times 32$ images with power spectral density proportional to $1/f^2$ (since the power spectra of natural images have been shown to follow a $1/f^\alpha$ trend, where $f$ is the spatial frequency in cycles per image~\citep{VANDERSCHAAF19962759}, and $\alpha$ is typically between $1$ and $3$). The CIFAR-10, Warped CIFAR-10, and $1/f^2$ priors were all trained with the same number of training images ($45$K), with the same tapering effect described in Section \ref{subsec:priors}, and for the same number of iterations ($100$K). \textbf{(a)} Nine training samples and the average log power spectrum of $10$K training samples (without the taper) for each prior. The average spectral power of CIFAR-10 images is relatively high in the horizontal and vertical frequencies. Warping the CIFAR-10 images seems to reduce the preference for vertical frequencies, but a large presence of horizontal frequencies remains (perhaps in part because horizons are still distinguishable after warping, as can be seen in some of the training samples). The $1/f^2$ noise images have isotropically-distributed spectral power. \textbf{(b)} Results of imaging the Crescent simulated data under the different priors. A posterior sample and the average of $128$ posterior samples are shown for each prior. With the regular CIFAR-10 prior, there is a sharp corner at the top-right of the ring-like object that makes it look squarish. This artifact is not present in the Warped CIFAR-10 reconstructions, which more resemble a smooth ring. The $1/f^2$ prior, which does not have a preference for any frequency orientation, also recovers an object that resembles a smooth ring.
\label{fig:boxyness}}
\end{figure*}

While CIFAR-10 represents a ``generic'' natural-image prior, the dataset itself still contains biases. The CIFAR-10 dataset comprises upright images of animals and man-made objects, which tend to exhibit horizontal or vertical lines. As Figure \ref{fig:boxyness}(a) shows, the average log power spectrum of CIFAR-10 images has most power in the purely horizontal or vertical spatial frequencies. This preference for vertically- or horizontally-oriented edges results in images of objects that look somewhat rectangular instead of circular, even given measurements of a ring structure. See, for example, the CIFAR-10 reconstructions from the Crescent and GRMHD 1 data in Figures \ref{fig:simdata} and \ref{fig:simdata_posterior} or the April 10 and 11 CIFAR-10 reconstructions of M87* in the following section. In Figure \ref{fig:boxyness}, we demonstrate on the Crescent data how boxy artifacts can be mitigated with a prior trained on warped CIFAR-10 images or a prior trained on images with a $1/f^2$ spectral distribution. By distorting CIFAR-10 images with warped random affine transforms, we expect to reduce the presence of straight lines by perturbing them to have more curvature. Alternatively, by randomly sampling from a $1/f^2$ spectral distribution, we create a dataset of images that follow a simplified statistical model without any preference for straight lines.

\newpage
\section{M87* Results}
\label{sec:m87_results}
We estimated image posteriors of M87* given the EHT observations from 2017 April. For each of the four observation days, we gathered the closure phases and log closure amplitudes from the combined low-band and high-band public data.\footnote{\url{https://datacommons.cyverse.org/browse/iplant/home/shared/commons_repo/curated/EHTC_FirstM87Results_Apr2019}} We used the same data preprocessing and visualization steps as for the synthetic data.

\begin{figure}[t]
    \centering
    \includegraphics[width=0.5\textwidth]{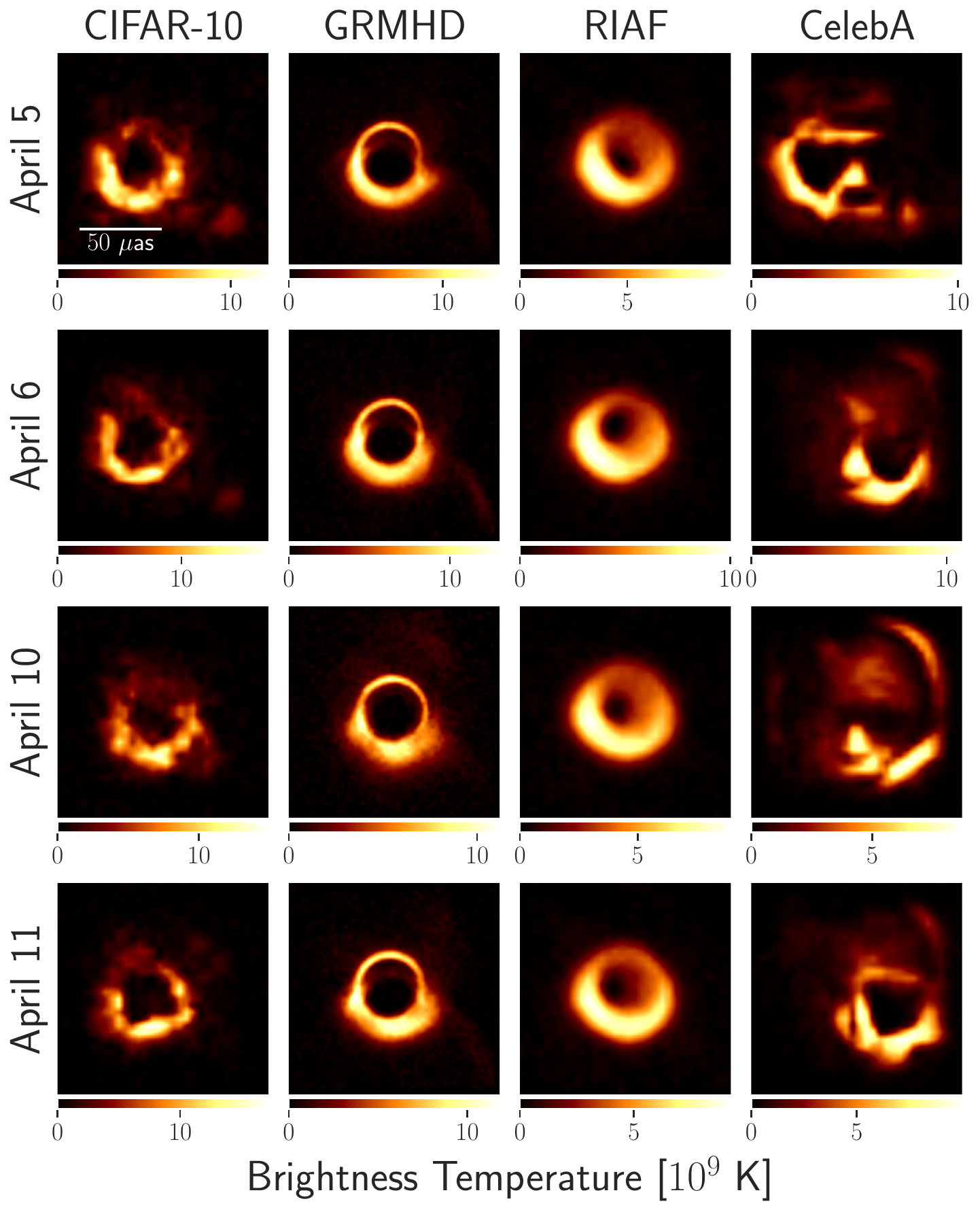}
    \caption{M87* posterior samples. A random posterior image is shown for each observation day and prior. The CIFAR-10 prior was trained on images of everyday objects and makes no assumptions of black-hole structure, yet it recovers a ring-like structure for all four observation days. The GRMHD prior assumes a fluid-flow model of black holes, which helps it recover visually-striking images of a thin ring with wisps. The RIAF prior assumes a simplified crescent model of black holes, which results in simplified crescent images of M87*. The CelebA prior was trained on images of human faces, so its preferred images are presumably far away from the true source image. Even with its incorrect and strong biases, the CelebA prior recovers a ring-like structure, here favoring an eye from the face prior to explain the ring. These images under various priors all fit the EHT observations but incorporate different visual biases.}
    \label{fig:m87}
\end{figure}

Figure \ref{fig:m87} shows a posterior sample for each of the observation days, under each prior. The CIFAR-10 and CelebA priors, which incorporate no assumptions about black holes, lead to images with a ring-like structure. The CelebA reconstructions include some face-like details, especially for April 10, the day with the fewest observations and thus the day whose image is least constrained by measurements. Nonetheless, the fact that both the black-hole-agnostic priors recover ring shapes is strong evidence of a ring-like structure in the measurements. It also reveals that basic natural-image statistics shared by CIFAR-10 and CelebA may be sufficient to retrieve the ring structure under a constrained FOV.

The GRMHD and RIAF priors lead to images with a well-defined ring structure. The GRMHD prior is visually richer, encouraging a thin ring with wispy features in the bright region of the ring. The RIAF prior constrains the image according to a simplified geometric model without adding any lower-level details.
% The GRMHD images are visually richer, and one may prefer them if one believes in the GRMHD model.

Table \ref{tab:m87_chisq}, which reports reduced $\chi^2$ values, confirms that all the reconstructed images are consistent with the EHT measurements. Across priors and observation days, the $\chi^2$ values are $<1.5$, which is considered a good sign of fitting the measured data. As with the simulated data, the RIAF prior struggles most to fit the data (most of its $\chi^2$ values are greater than $1$) because it imposes the most constraining black-hole model. The rest of the priors tend to result in $\chi^2<1$, which means that they are flexible enough to somewhat overfit the data and that any differences between their posteriors are likely due to the visual biases of the priors that are not constrained by the data.

\begin{deluxetable*}{rLCCCC}
\label{tab:m87_chisq}
\tablewidth{0pt}
\tablehead{
\nocolhead{Image} & \nocolhead{Data Product} & \colhead{CIFAR-10} & \colhead{GRMHD} & \colhead{RIAF} & \colhead{CelebA}
}
\startdata
April 5 & \chi^2_\mathrm{cp} & 0.80 \pm 0.03 & 0.83 \pm 0.15 & 1.27 \pm 0.06 & 0.80 \pm 0.03 \\ 
& \chi^2_\mathrm{logca} & 0.75 \pm 0.04 & 0.76 \pm 0.09 & 1.50 \pm 0.10 & 0.71 \pm 0.04 \\ \hline
April 6 & \chi^2_\mathrm{cp} & 0.92 \pm 0.02 & 0.93 \pm 0.05 & 1.08 \pm 0.03 & 0.93 \pm 0.03 \\ 
& \chi^2_\mathrm{logca} & 0.76 \pm 0.02 & 0.74 \pm 0.04 & 1.00 \pm 0.05 & 0.80 \pm 0.06 \\ \hline
April 10 & \chi^2_\mathrm{cp} & 0.92 \pm 0.05 & 0.89 \pm 0.04 & 1.25 \pm 0.07 & 0.90 \pm 0.05 \\ 
& \chi^2_\mathrm{logca} & 0.73 \pm 0.06 & 0.71 \pm 0.04 & 1.35 \pm 0.12 & 0.72 \pm 0.06 \\ \hline
April 11 & \chi^2_\mathrm{cp} & 1.09 \pm 0.02 & 1.06 \pm 0.02 & 1.26 \pm 0.04 & 1.08 \pm 0.03 \\ 
& \chi^2_\mathrm{logca} & 0.69 \pm 0.04 & 0.65 \pm 0.04 & 1.01 \pm 0.06 & 0.65 \pm 0.03
\enddata
\caption{Data-consistency metrics ($\chi^2$) for closure quantities of M87* data. $\chi_\mathrm{cp}$ and $\chi_\mathrm{logca}$ are the $\chi^2$ metrics for closure phases and log closure amplitudes, respectively. The avg.~$\pm$ std.~dev.~of $128$ samples from the estimated posterior is reported. $\chi^2\approx 1$ indicates a good balance between fitting the observed data and fitting the prior. Lower $\chi^2$ means more data consistency. The RIAF prior leads to the highest $\chi^2$ values, meaning the strength of this prior relative to the data is strongest. This is probably because the RIAF model is the most constraining of the priors.}
\end{deluxetable*}

\subsection{Characterizing the Posteriors}
\begin{figure}[ht]
    \centering
    \includegraphics[width=0.85\textwidth]{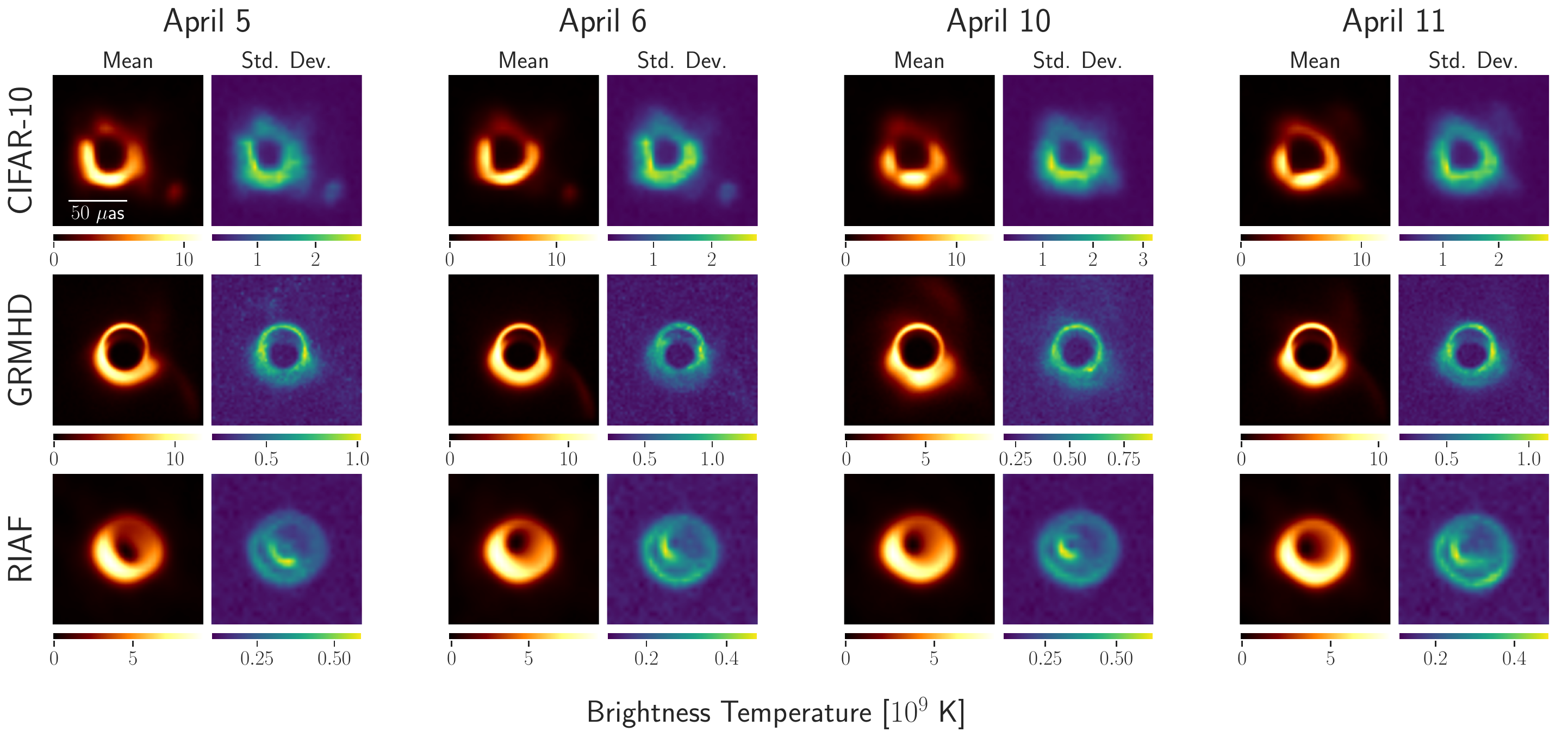}
    \caption{Mean and std.~dev.~of posterior samples. Like with the simulated data in Figure \ref{fig:simdata_posterior}, uncertainty decreases as the prior becomes stronger. Note that under the CIFAR-10 prior the day with the most uncertainty is April 10, the day with the least data. The stronger priors (i.e., GRMHD and RIAF) do not exhibit more uncertainty for this day compared to other days.}
    \label{fig:m87_posterior}
\end{figure}

Our imaging approach leads to single-mode M87* image posteriors, except for some CelebA posteriors that are bimodal. Figure \ref{fig:m87_posterior} shows Gaussian fits of the posteriors estimated under the CIFAR-10, GRMHD, and RIAF priors. Under all these priors, the mean image shows a clear progression of the bright spot of the ring moving counter-clockwise over the four observation days. The pixel-wise standard deviation shows areas of uncertainty around the mean. Similarly to our results with simulated data, the results in Figure \ref{fig:m87_posterior} show that as the prior becomes stronger (i.e., from CIFAR-10 to GRMHD to RIAF), the uncertainty goes down. Under the CIFAR-10 and GRMHD priors there is uncertainty throughout the ring, whereas under the RIAF prior uncertainty lies mainly along the edges of the ring.

\begin{figure}[ht]
    \centering
    \includegraphics[width=0.8\textwidth]{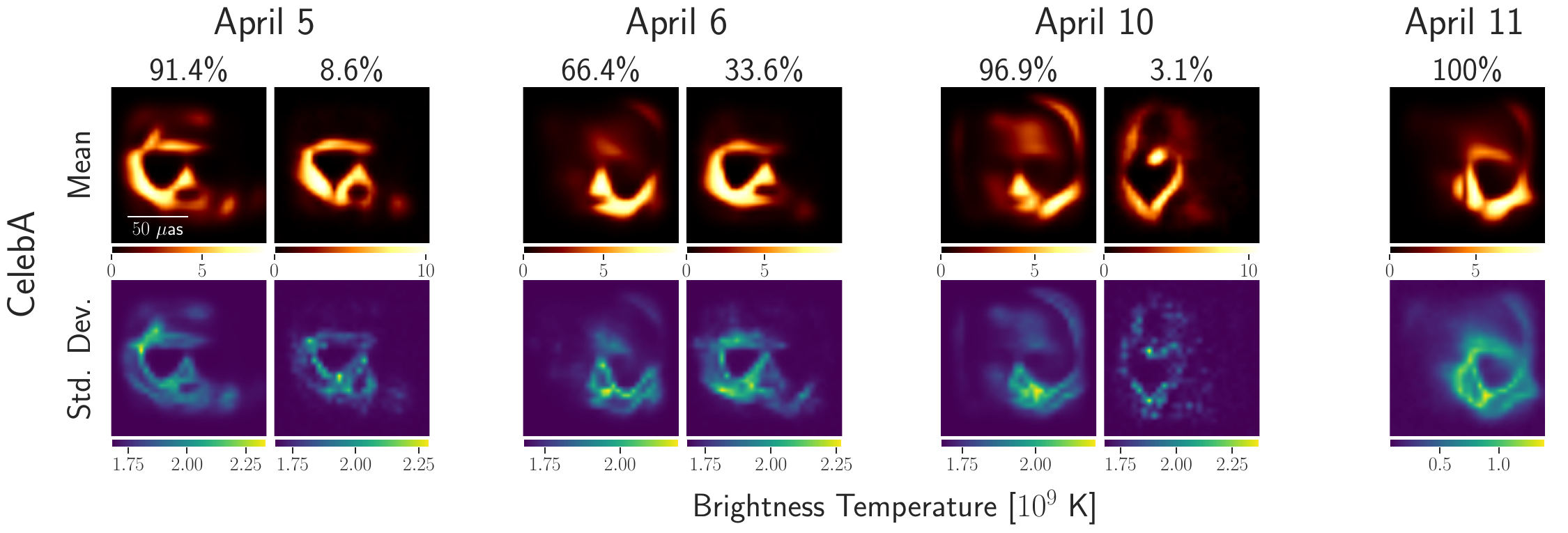}
    \caption{Bimodal M87* image posteriors under the CelebA prior. A two-component Gaussian mixture model was fit to $128$ posterior samples for April 5, 6, and 10. The mean, std.~dev., and weight of each Gaussian mixture component are shown. The only single-mode posterior is for April 11, which is the day with the most $(u,v)$ coverage.}
    \label{fig:m87_celeb}
\end{figure}

Figure \ref{fig:m87_celeb} visualizes the bimodal posteriors under the CelebA prior. Like with CIFAR-10, the magnitude of the uncertainty is higher than that of the GRMHD and RIAF priors. The presence of multiple modes further reflects higher uncertainty under the CelebA prior, which makes sense for a prior that is so mismatched with the observed data.

A feature that stands out in the CIFAR-10, GRMHD, and CelebA reconstructions is a southwest region of extended flux outside the ring. It is especially noticeable in the April 5 and 6 images. See, for example, the faint spot of brightness in the CIFAR-10 reconstructions and the faint wisp to the southwest in the GRMHD reconstructions in Figures \ref{fig:m87} and \ref{fig:m87_posterior}. A disconnected southwest region of brightness also appears in both modes of the CelebA posterior on April 5 and in the second mode of the April 6 posterior in Figure \ref{fig:m87_celeb}. Such a feature is not visible in the RIAF reconstructions. With previous imaging results that only incorporated one prior, it would have been difficult to conclude whether this feature was an artifact of imaging or a clue from the data. Our results in Figures \ref{fig:m87_posterior} and \ref{fig:m87_celeb} suggest that it is a prior-dependent feature, with different priors placing different amounts of brightness in that southwest region.

To summarize our findings from the estimated M87* posteriors, the most notable result is that all priors recover ring-like structure. The priors that do not assume a black hole (i.e., CIFAR-10 and CelebA) exhibit most uncertainty in the posterior and are most flexible with adding flux outside of the ring. The priors based on a black-hole model (i.e., GRMHD and RIAF) reconstruct the clearest rings, with the GRMHD prior providing the most visual detail. In general, the more constraining the prior, the less uncertainty there is in the posterior, but the more potential there is to overfit to prior assumptions.

\section{Extracted Ring Features}
\label{sec:features}
An important stage of analysis is to extract ring features from reconstructed images, distilling any ring structure into a few parameters dictating its geometry and brightness profile.
In \citet{m87paperiv,m87papervi}, the EHT Collaboration extracted, among other quantities, the following characteristic ring features: diameter, width, orientation, asymmetry, and fractional central brightness. It found the ring diameter and orientation angle to be most consistent across imaging methods, with the other quantities varying depending on the imaging pipeline. In this paper, we focus on these characteristic features and analyze the effect of the prior on them.

\begin{figure}
    \centering
    \includegraphics[width=0.7\textwidth]{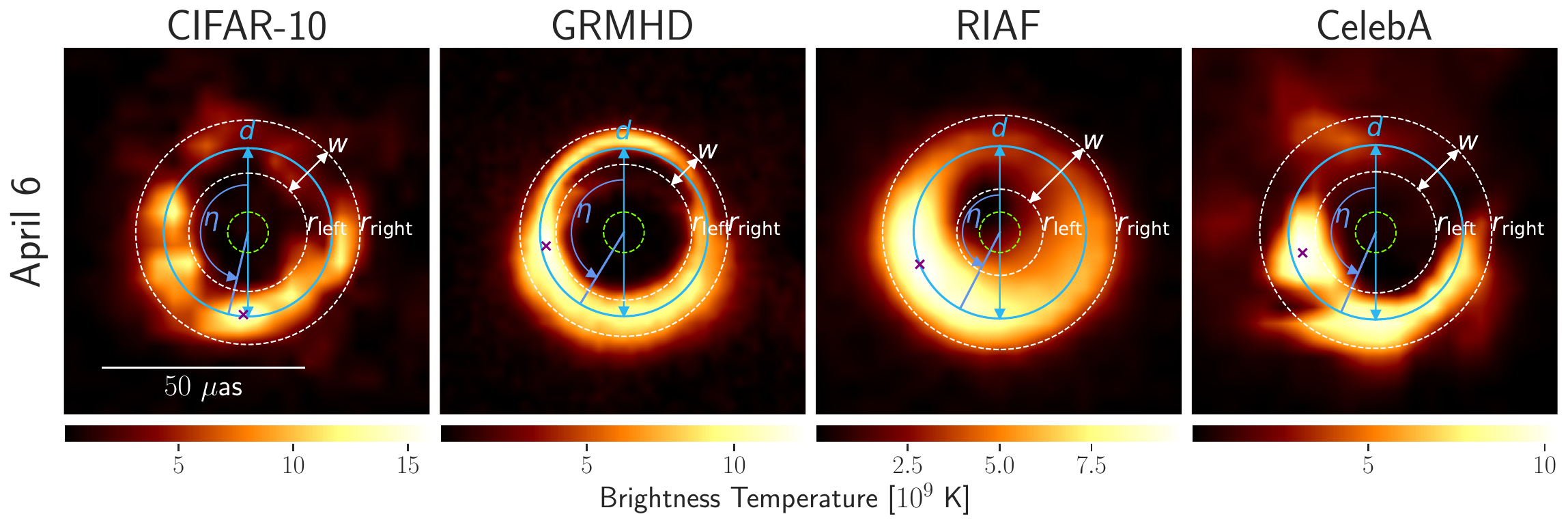}
    \caption{Visualization of extracted ring features of M87* images. Shown is a random sample (centered by \texttt{REx}) from each posterior for April 6. Parameter $d$ is the ring diameter; $w$ is the ring width. Parameters $r_\text{left}$ and $r_\text{right}$ delimit the radial FWHM used to estimate the orientation angle and asymmetry. Parameter $\eta$ is the orientation angle measured east of north. The purple cross marks the location of peak brightness. Asymmetry $A$ and fractional central brightness $f_\text{C}$ are not visualized. The green dashed circle demarcates the inner disk of radius $5$ $\mu$as used to define $f_\text{C}$.}
    \label{fig:m87_ring_fits}
\end{figure}

The \textbf{diameter} $d$ indicates the full size of the ring and is calculated based on the distance between the peak brightness and the ring center. The \textbf{width} $w$ indicates the thickness of the ring itself. The \textbf{orientation} angle $n$, measured east of north, roughly indicates the radial position of most of the brightness. The azimuthal \textbf{asymmetry} $A$, a measure of brightness asymmetry throughout the ring, roughly indicates the magnitude of brightness at the measured orientation. The \textbf{fractional central brightness} $f_\mathrm{C}$ indicates how bright the interior of the ring is compared to the ring itself and can be considered an inverse brightness contrast ratio. We use the \texttt{REx} feature extraction algorithm~\citep{chael2019simulating} implemented in the \texttt{eht-imaging} library. $d$ and $f_\text{C}$ are measured the same way as in \citet{m87paperiv}, but the other features have slight differences (we default to following the \texttt{REx} implementation rather than the equations provided in \citet{m87paperiv}). Appendix~\ref{app:feature_extraction} contains the exact equations we used to compute these features. Figure \ref{fig:m87_ring_fits} helps visualize some of the features.

\subsection{Ring Features of Simulated-data Reconstructions}
\begin{figure}[t]
    \centering
    \includegraphics[width=0.9\textwidth]{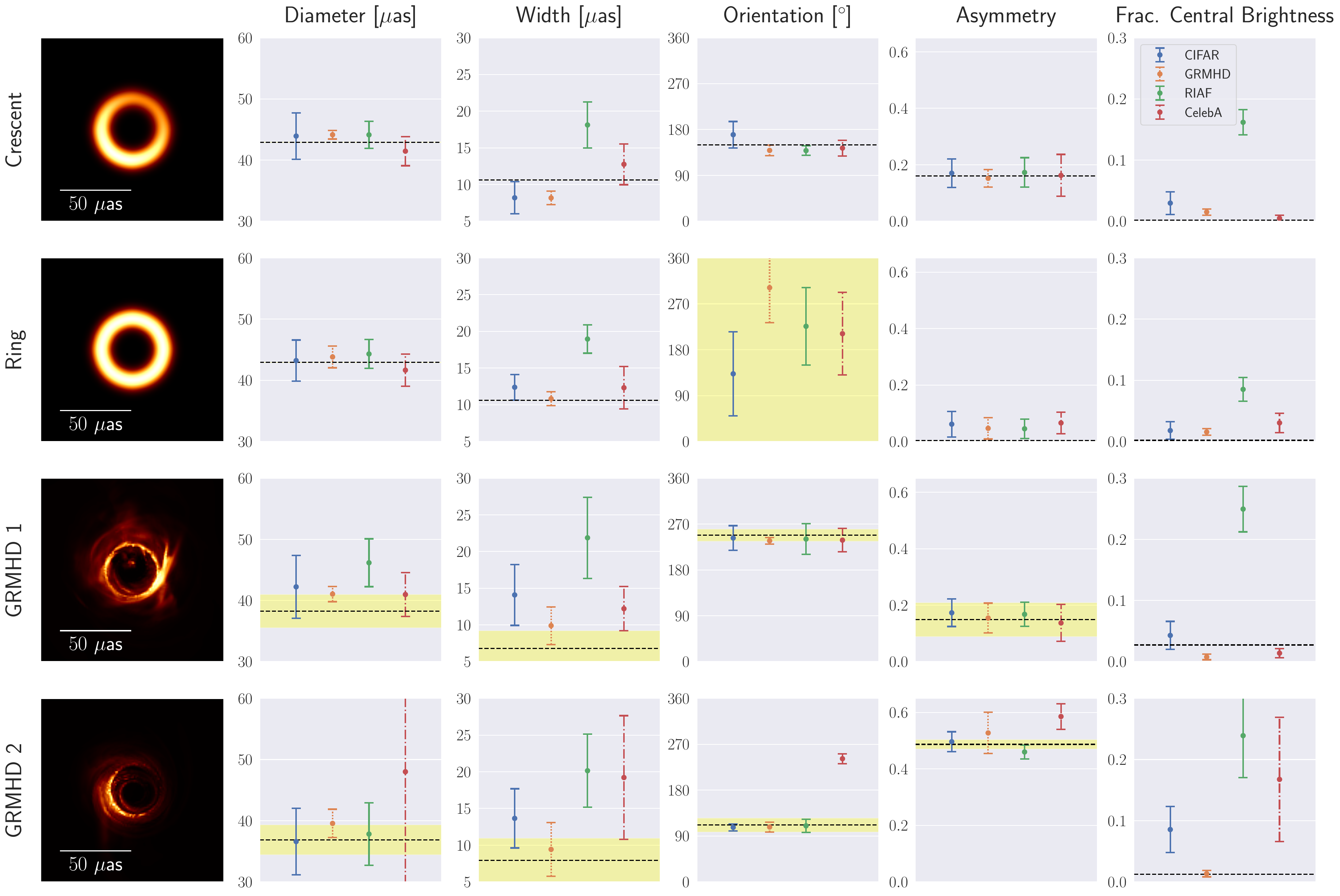}
    \caption{Extracted ring features of images reconstructed from simulated data. The dotted line indicates the measured quantity of the ground-truth image, and the shaded yellow region indicates the std.~dev.~as computed by REx (the orientation of the Ring image was manually set to span $0^\circ$--$360^\circ$). For the score-based priors (CIFAR, GRMHD, RIAF, CelebA), the mean was computed across the measured quantities of $128$ posterior samples. The std.~dev.~for all features except fractional central brightness ($f_\mathrm{C}$) was computed as the square root of the average variance across these samples. Since REx does not provide uncertainty for $f_\text{C}$ of a single image, the $f_\text{C}$ error bars were computed as the std.~dev.~of the $f_\text{C}$ values of all posterior samples. All choices of prior roughly recover the target diameter, orientation, and asymmetry. The only exceptions are that the RIAF prior over-estimates the diameter of GRMHD 1 (due to the difficulty of fitting GRMHD data to a strong RIAF prior) and that the CelebA prior over-biases the orientation of GRMHD 2 (due to the CelebA prior struggling to produce a strong ring structure with such data). The width and fractional central brightness are particularly prior-dependent, with the GRMHD prior providing consistently small values for these features. In contrast, the RIAF prior prefers large widths and high $f_\text{C}$ values. Please see Figure \ref{fig:simdata} for image samples.}
    \label{fig:simdata_features}
\end{figure}

We performed feature extraction on images from simulated data, focusing on the source images with ring-like structure: Crescent, Ring, GRMHD 1, and GRMHD 2. We analyzed the recovered posteriors under all the score-based priors but excluded the non-ring-like samples from the CelebA posteriors for the Ring and GRMHD 1 data (see Figure \ref{fig:simdata}, which shows that the second mode of the Ring posterior and the first mode of the GRMHD 1 posterior do not have a ring-like structure). Figure \ref{fig:simdata_features} shows all the extracted features and their error bars. In the following paragraphs, we discuss each feature and its dependence (or lack thereof) on the prior.

\paragraph{Diameter}
All priors recover the diameter of the source image (within one standard deviation from the mean extracted diameter). The only exception is that the RIAF prior leads to a larger diameter for GRMHD 1. A possible explanation is that the RIAF model is too strong of a prior for these data, as it must account for all the flux in the image with a thick crescent (as evidenced by the relatively high extracted diameter and width of the RIAF-reconstructed images). The RIAF prior also leads to the highest $\chi^2$ values (Table \ref{tab:simdata_chisq}), further evidence that it has difficulty fitting the GRMHD 1 observations. On the other hand, the RIAF prior correctly recovers the diameter and has lower $\chi^2$ values for the Ring and Crescent, two objects that can be well-approximated with a RIAF model. Even though the diameter should be well-constrained by the measurements, our results demonstrate how a strong-enough prior may recover a ring structure but with an incorrect diameter.

The CIFAR-10 prior gives the most accurate mean diameter, although with greater uncertainty than the GRMHD and RIAF priors. This makes sense, as the CIFAR-10 prior, in making the weakest assumptions, is most flexible with the image reconstruction. The CelebA prior exhibits significantly high uncertainty for GRMHD 2, perhaps because its recovered images have the least ring-like structure (Figure \ref{fig:simdata}). The GRMHD prior accurately recovers the diameter across all these ring-like data and with relatively little uncertainty.

\paragraph{Width} The ring width varies significantly with the prior, which supports previous findings that the width is less well-constrained by the measurements than the diameter is \citep{m87paperiv,m87papervi}. The GRMHD prior recovers the thinnest rings as a result of being trained on GRMHD images that exhibit thin rings.

\paragraph{Orientation} All priors recover the orientation angle of the source image, except the CelebA prior given GRMHD 2 data. Like the high diameter uncertainty, this may be due to the reconstructed images having relatively weak ring structure. As can be seen in the two samples from this posterior in Figure \ref{fig:simdata}, there is actually some brightness outside of the ring-like area. These results further highlight that strong and incorrect assumptions in the prior may inhibit correct recovery of ring features that should be constrained by the observations. Besides this extreme case of applying a CelebA prior to GRMHD data, the ring orientation appears to be independent of the prior.

\paragraph{Asymmetry} The brightness asymmetry appears to be fairly robust to the prior. Again, the exception is the CelebA prior applied to the GRMHD 2 data, which can be explained by the weak ring structure present in the CelebA-recovered images. We note that for the Ring data, all priors produce a slight asymmetry even though the ground-truth object has no asymmetry, which is unavoidable with most imaging algorithms~\citep{m87paperiv}.

\paragraph{Fractional central brightness} This feature varies extremely with the prior, and the original M87* imaging work also found that $f_\text{C}$ is not well-constrained by the data~\citep{m87paperiv}. The GRMHD prior recovers the lowest fractional central brightness for the GRMHD observations. Along with smaller ring widths, these low $f_\text{C}$ values indicate that using a well-matched GRMHD prior on GRMHD observations gives the benefit of sharper images than could be obtained with a weaker prior like CIFAR-10.

\subsubsection{Discussion}
To summarize our results on simulated data, we find that the recovered ring diameter, orientation angle, and asymmetry are fairly robust to the image prior, while the width and fractional central brightness are tied to the prior. However, it is possible to overly bias the diameter with an overly-biased prior. In particular, the RIAF prior applied to observations of a GRMHD simulation with flux extending beyond the ring may be too constraining and cause over-estimation of the diameter. Reassuringly, priors that are less constraining or more accurate dependably recover the diameter, orientation, and asymmetry. On the other hand, features like the ring width and fractional central brightness can be adjusted by imposing different priors. 
% The GRMHD prior generally recovers crisp images of thin rings with high contrast. It is important to remember that such prior-dependent features are not grounded in the data and only improve the visual experience of the images.

\subsection{Ring Features of M87* Reconstructions}
\begin{figure}[t]
    \centering
    \includegraphics[width=0.9\textwidth]{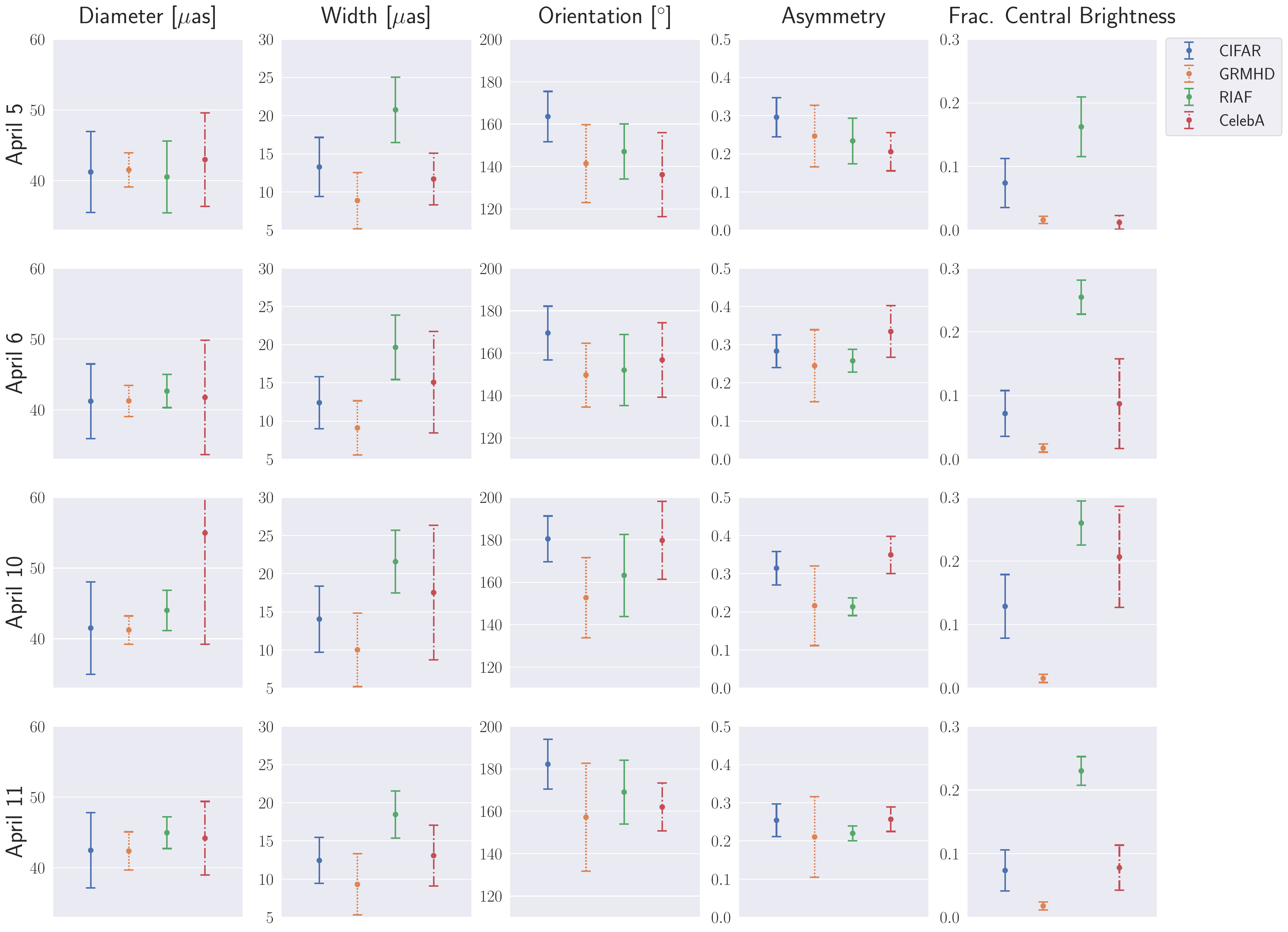}
    \caption{Extracted ring features of M87* images. The means and error bars were computed the same way as for Figure \ref{fig:simdata_features}. The different priors (CIFAR, GRMHD, RIAF, CelebA) all agree on the diameter, orientation, and asymmetry up to error bars. There is some disagreement in asymmetry for April 10, which is the day with fewest observations. We note a slight upward trend in diameter and orientation angle over the observation days. The width and fractional central brightness change with the prior, with the GRMHD prior providing the thinnest rings and most brightness contrast. Please see Figure \ref{fig:m87} for image samples.}
    \label{fig:m87_features}
\end{figure}

Figure \ref{fig:m87_features} and Table \ref{tab:m87_features} show the results of feature extraction for the M87* image reconstructions. Table \ref{tab:m87_features} includes the results of using the \texttt{eht-imaging} algorithm with fiducial parameters~\citep{m87paperiv} for reference, although \texttt{eht-imaging} is not directly comparable to our score-based priors since it utilizes hand-crafted regularizers as a proxy for a prior and visibility amplitudes instead of closure amplitudes. Figure \ref{fig:m87_ring_fits} visualizes the ring features on April 6 reconstructions.
We assumed ring structure in all the posterior samples under the different score-based priors. As shown in Figures \ref{fig:m87} and \ref{fig:m87_celeb}, the CelebA images have the weakest ring structure, which may have caused higher variance in the extracted features.

The CIFAR-10 prior recovers a mean diameter of $41.2$--$42.5$ $\mu$as across the four observations days. Accounting for error bars, all score-based priors agree on a range of possible diameters. The diameters recovered by our score-based priors are consistent with the diameter recovered by \texttt{eht-imaging}, although with slightly higher means and larger error bars. Like with the simulated data, the RIAF prior leads to relatively high diameters. Combined with the fact that the RIAF prior has the highest $\chi^2$ values in Table \ref{tab:m87_chisq}, this suggests that the RIAF model is likely too constraining for M87*.

The ring width depends on the prior, with the GRMHD prior resulting in the lowest width (about $9$ $\mu$as). The RIAF prior causes the largest width (about $20$ $\mu$as), which is only slightly larger than the width recovered by \texttt{eht-imaging} (about $16$ $\mu$as, as listed in Table \ref{tab:m87_features}).

The orientation angle is consistent across the priors, roughly ranging from $150^\circ$ on April 5 to $170^\circ$ on April 11. Like previous work, we find that both the diameter and orientation angle have an upward trend over the observation days as brightness shifts to the southwest \citep{m87paperiv,m87papervi}.

The amount of asymmetry is also roughly consistent across the priors. Interestingly, there is most discrepancy between the priors in the April 10 reconstructions. April 10 is the day with fewest observations, which may cause the brightness asymmetry to be more flexible under the data.

As expected, the fractional central brightness varies significantly with the prior. As with the simulated data, the GRMHD prior achieves the greatest brightness contrast.

Overall, the conclusions regarding the effects of priors on the M87* reconstructions are consistent with the conclusions of previous analyses \citep{m87paperiv,m87papervi} and our simulated-data analysis. The diameter, orientation, and asymmetry are robust to the prior, although some priors result in greater variability than others. There is a slight upward bias in diameter from the RIAF prior, due to wider rings that account for all the flux in one RIAF model. The most prior-dependent features are the width and fractional central brightness.

\begin{deluxetable*}{rLCCCC}
\label{tab:m87_features}
\tablewidth{0pt}
\tablehead{
\nocolhead{Image} & \colhead{\texttt{eht-imaging}*} & \colhead{CIFAR-10} & \colhead{GRMHD} & \colhead{RIAF} & \colhead{CelebA}
}
\startdata
\multicolumn{1}{l}{Diameter [$\mu$as]} \\ 
April 5 & 39.3 \pm 1.6 & 41.2 \pm 5.7 & 41.5 \pm 2.4 & 40.5 \pm 5.1 & 43.0 \pm 6.6 \\ 
April 6 & 39.5 \pm 1.5 & 41.2 \pm 5.3 & 41.2 \pm 2.2 & 42.6 \pm 2.3 & 41.7 \pm 8.1 \\ 
April 10 & 40.5 \pm 1.3 & 41.5 \pm 6.5 & 41.2 \pm 2.0 & 44.0 \pm 2.8 & 55.0 \pm 15.8 \\ 
April 11 & 41.1 \pm 1.2 & 42.5 \pm 5.3 & 42.4 \pm 2.7 & 45.0 \pm 2.2 & 44.2 \pm 5.2 \\ \hline 
\multicolumn{1}{l}{Width [$\mu$as]} \\ 
April 5 & 16.3 \pm 1.5 & 13.3 \pm 3.9 & 8.9 \pm 3.7 & 20.8 \pm 4.3 & 11.7 \pm 3.4 \\ 
April 6 & 16.2 \pm 1.0 & 12.4 \pm 3.4 & 9.1 \pm 3.5 & 19.7 \pm 4.2 & 15.1 \pm 6.6 \\ 
April 10 & 15.7 \pm 1.3 & 14.1 \pm 4.3 & 10.0 \pm 4.8 & 21.6 \pm 4.1 & 17.5 \pm 8.8 \\ 
April 11 & 15.6 \pm 0.9 & 12.4 \pm 3.0 & 9.3 \pm 4.0 & 18.5 \pm 3.1 & 13.1 \pm 4.0 \\ \hline 
\multicolumn{1}{l}{Orientation [$^\circ$]} \\ 
April 5 & 149.0 \pm 4.0 & 163.5 \pm 11.9 & 141.4 \pm 18.4 & 147.0 \pm 13.0 & 136.2 \pm 19.9 \\ 
April 6 & 151.2 \pm 3.2 & 169.5 \pm 12.6 & 149.7 \pm 15.1 & 152.0 \pm 16.8 & 156.8 \pm 17.5 \\ 
April 10 & 171.1 \pm 3.4 & 180.4 \pm 10.8 & 152.7 \pm 18.9 & 163.2 \pm 19.4 & 179.7 \pm 18.4 \\ 
April 11 & 167.5 \pm 3.1 & 182.2 \pm 11.7 & 157.2 \pm 25.5 & 169.0 \pm 15.1 & 162.0 \pm 11.3 \\ \hline 
\multicolumn{1}{l}{Asymmetry} \\ 
April 5 & 0.25 \pm 0.01 & 0.30 \pm 0.05 & 0.25 \pm 0.08 & 0.23 \pm 0.06 & 0.21 \pm 0.05 \\ 
April 6 & 0.24 \pm 0.02 & 0.28 \pm 0.04 & 0.24 \pm 0.09 & 0.26 \pm 0.03 & 0.33 \pm 0.07 \\ 
April 10 & 0.23 \pm 0.00 & 0.31 \pm 0.04 & 0.22 \pm 0.10 & 0.21 \pm 0.02 & 0.35 \pm 0.05 \\ 
April 11 & 0.20 \pm 0.01 & 0.25 \pm 0.04 & 0.21 \pm 0.11 & 0.22 \pm 0.02 & 0.26 \pm 0.03 \\ \hline 
\multicolumn{1}{l}{Frac. Central Brightness} \\ 
April 5 & 0.07 & 0.07 \pm 0.04 & 0.02 \pm 0.01 & 0.16 \pm 0.05 & 0.01 \pm 0.01 \\ 
April 6 & 0.07 & 0.07 \pm 0.04 & 0.02 \pm 0.01 & 0.25 \pm 0.03 & 0.09 \pm 0.07 \\ 
April 10 & 0.04 & 0.13 \pm 0.05 & 0.02 \pm 0.01 & 0.26 \pm 0.03 & 0.21 \pm 0.08 \\ 
April 11 & 0.04 & 0.07 \pm 0.03 & 0.02 \pm 0.01 & 0.23 \pm 0.02 & 0.08 \pm 0.04 \\ 
\enddata
\caption{Extracted ring features of M87* images. This table reports the means and std.~devs.~that are visualized in Figure \ref{fig:m87_features}. *\texttt{eht-imaging} is not exactly comparable to the score-based priors because it is a different imaging algorithm that uses hand-crafted regularizers instead of a data-driven prior. There are small differences from the values for \texttt{eht-imaging} reported in Table 7 of \citet{m87paperiv}. This may be due to differences in implementation/hardware and in the exact definitions of the features (see Appendix \ref{app:feature_extraction} for details).}
\end{deluxetable*}

\section{Discussion}
\label{sec:discussion}
We have presented a strategy for probing the influence of different priors on imaging from EHT VLBI data. Our imaging approach allows one to infer a diverse collection of image posteriors, each incorporating different assumptions. From these results, one can analyze the effect of the prior on the visual quality and uncertainty of reconstructed images, as well as on the recovered ring features. We applied this strategy to EHT observations of M87* and found that a GRMHD prior is most effective at recovering a visually-detailed image with a thin ring and sharp contrast. However, it exhibits less posterior uncertainty than some other priors do, meaning that one should only trust the GRMHD reconstructions if one believes in the existence of an underlying compact ring structure. On the other hand, it is possible to allow for more flexibility in the posterior with a weaker prior. For example, a CIFAR-10 prior still recovers a ring with similar diameter, orientation, and brightness asymmetry, albeit with blurrier quality. Our proposed strategy allows scientists to reliably interpret reconstructed images and account for the role of prior assumptions in their analysis.

\section*{Acknowledgments}
% \begin{acknowledgments}
This work was funded by NSF 2048237, 2034306, 1935980,
1955864, and PHY-2019786 and the Amazon AI4Science
Partnership Discovery Grant. B.T.F. is supported by the NSF GRFP (NSF Grant No.~2139433).
The authors thank Charles Gammie, Ben Prather, Abhishek Joshi, Vedant Dhruv, and C. K. Chan for sharing their GRMHD simulations, as well as Aviad Levis for many helpful discussions.
% \end{acknowledgments}

\newpage
\appendix

\counterwithin*{equation}{section}

\section{Flux Constraint and Data Annealing}
\label{app:ablation}
\begin{figure}[ht]
    \centering
    \includegraphics[width=0.95\textwidth]{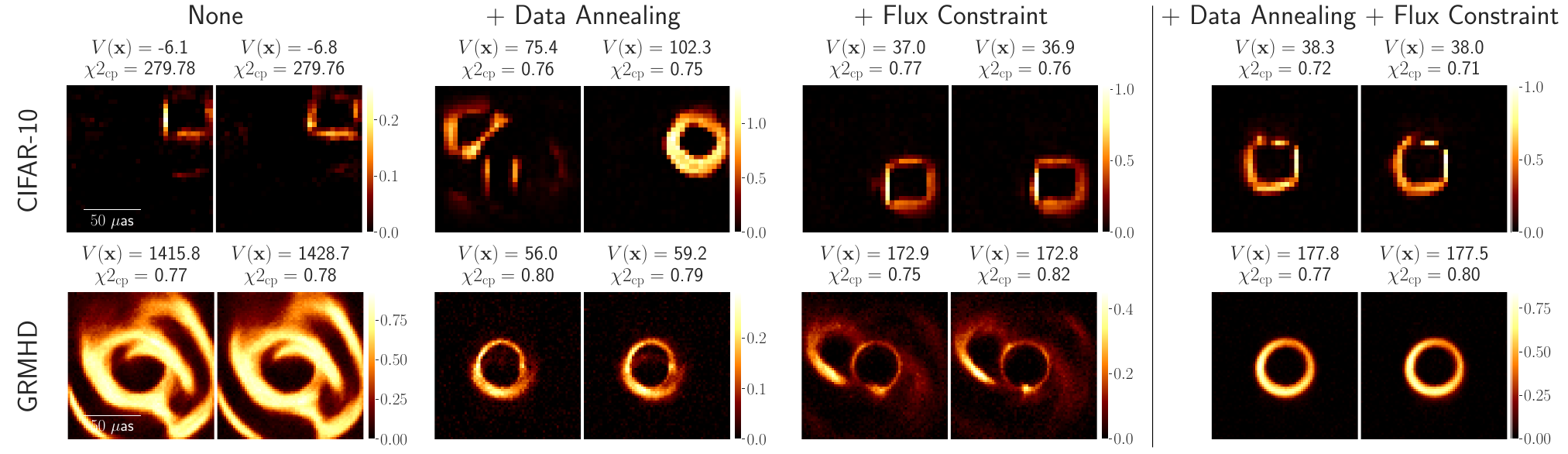}
    \caption{Ablation of data annealing and flux constraint. The simulated measurements are of the Crescent image in Figure \ref{fig:simdata}. The posterior samples are shown with their original number of pixels (i.e., $32\times 32$ or $64\times 64$) and with their original pixel values (which should be between 0 and 1). ``None'' is vanilla optimization without data-weight annealing or a flux-constraint objective. Without a flux constraint, the total flux $V(\mathbf{x})$ of samples varies widely; furthermore, a posterior that is unconstrained in the total flux may be too complicated to model with VI. Without data annealing, optimization may become unstable, as is the case for the GRMHD prior here. Optimization is generally more stable when both data annealing and a flux constraint are used.}
    \label{fig:ablation}
\end{figure}

As mentioned in Section \ref{sec:method}, we make two additions to a vanilla VI approach: a flux-constraint objective and a data-weight annealing schedule. Figure \ref{fig:ablation} demonstrates the effect of each addition through an ablation study. A flux constraint imposes the assumption that all posterior images have the same total flux, which reduces the complexity of the posterior distribution. Data annealing helps prevent falling into bad local minima when optimizing the variational distribution (Equation \eqref{eq:objective}), since the magnitude of the data loss is much higher than the magnitude of the prior loss at the beginning of optimization. As Figure \ref{fig:ablation} shows, both additions help make optimization more stable.

\subsection{Flux Constraint}
\label{sec:flux}
Recall that a flux constraint is imposed through the following objective term in Equation \eqref{eq:objective}:
\begin{align}
    \mathcal{L}_\text{flux}(\mathbf{x})=\left(V(\mathbf{x})-\bar{V}\right)^2,
\end{align}
where $V(\mathbf{x})$ is the total flux of the image $\mathbf{x}$ and $\bar{V}$ is the target total flux.

\begin{figure}[ht]
    \centering
    \includegraphics[width=0.7\textwidth]{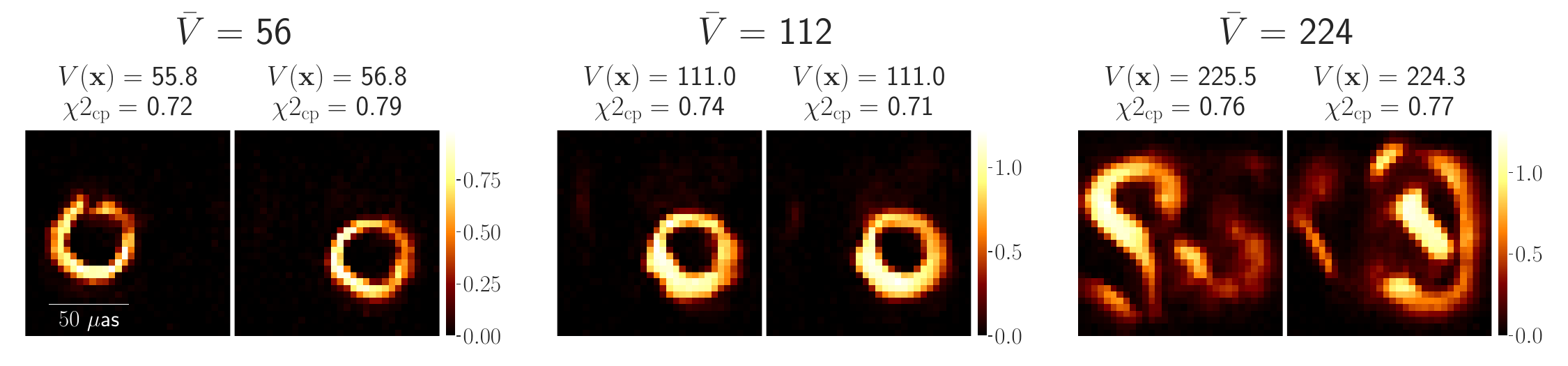}
    \caption{Effect of target flux $\bar{V}$. The simulated measurements are of the Crescent image in Figure \ref{fig:simdata}, and the prior is the CelebA score-based prior. $\bar{V}=112$ is the default target total flux chosen based on samples from the prior. When $\bar{V}$ is halved to $56$, the estimated posterior images are sparser. When $\bar{V}$ is doubled to $224$, the estimated posterior images have less compactness. $\bar{V}=112$ is essentially the preferred total flux under the score-based prior.}
    \label{fig:flux}
\end{figure}

The flux constraint should be considered an additional prior on top of the score-based prior. It imposes the assumption of a constant total flux across the posterior (since closure quantities in the log likelihood do not constrain total flux). The value of $\bar{V}$ also influences the posterior: a lower $\bar{V}$ encourages sparser, more-compact images than a higher $\bar{V}$. Figure \ref{fig:flux} demonstrates this when using the CelebA prior on simulated data of the Crescent image.

Our approach is to set $\bar{V}$ according to the preferred total flux of the score-based prior. That is, we set it as the median total flux of $512$ samples from the prior. This results in $\bar{V}=38,173,90,$ and $112$ for the CIFAR-10, GRMHD, RIAF, and CelebA priors, respectively. Images can be scaled to the actual target flux $\bar{V}_\text{orig}$ measured by the zero-baseline (e.g., $\bar{V}_\text{orig}=0.6$ Jy) by a factor of $\bar{V}_\text{orig}/\bar{V}$. By setting $\bar{V}$ based on the score-based prior, we attempt to make most of the image-reconstruction bias come from the score-based prior rather than from the target flux. 

\subsection{Data Annealing}
\label{app:data_annealing}
The variational distribution is randomly initialized to solve the VI optimization problem (Equation \eqref{eq:objective}), which means that the data loss, or negative log likelihood, is quite high at the beginning of optimization. This can cause convergence to posterior samples that are not preferred by the score-based prior (see the GRMHD example in Figure \ref{fig:ablation} when only a flux constraint is used). To avoid poor local minima, we initialize the weight of the data loss at $0$ and then gradually anneal it to $1$ with the following annealing schedule:
\begin{align}
    \lambda_\text{data}(i)=\frac{1}{1+\mathrm{exp}(-r(i-s))},
\end{align}
where $i\in\mathbb{Z}^+$ is the optimization step, $r\in[0,1]$ is the annealing rate, and $s\in\mathbb{Z}^+$ is the pivot step at which $\lambda_\mathrm{data}(i)=0.5$. We use $r=0.002$ and $s=12000$. Therefore, strictly speaking, our optimization approach is to iteratively apply the gradient of the following loss function with respect to the RealNVP parameters $\phi$ at each step $i$:
\begin{align}
\mathcal{L}(\phi,i) := \mathbb{E}_{\mathbf{x}\sim q_\phi}\left[\lambda_\text{data}(i)\left(-\log p(\mathbf{y}_\text{cp}\mid\mathbf{x})-\log p(\mathbf{y}_\text{logca}\mid\mathbf{x})\right)-b_\theta(\mathbf{x})+\left(V(\mathbf{x})-\bar{V}\right)^2+\log q_\phi(\mathbf{x})\right].
\end{align}
Since we run optimization for $100$K steps, $\lambda_\text{data}(i)$ is mostly equal to $1$ except at the very beginning of optimization.

\section{Model Uncertainty of the Prior}
\label{app:prior_uncertainty}
\begin{figure}[ht]
    \centering
    \includegraphics[width=\textwidth]{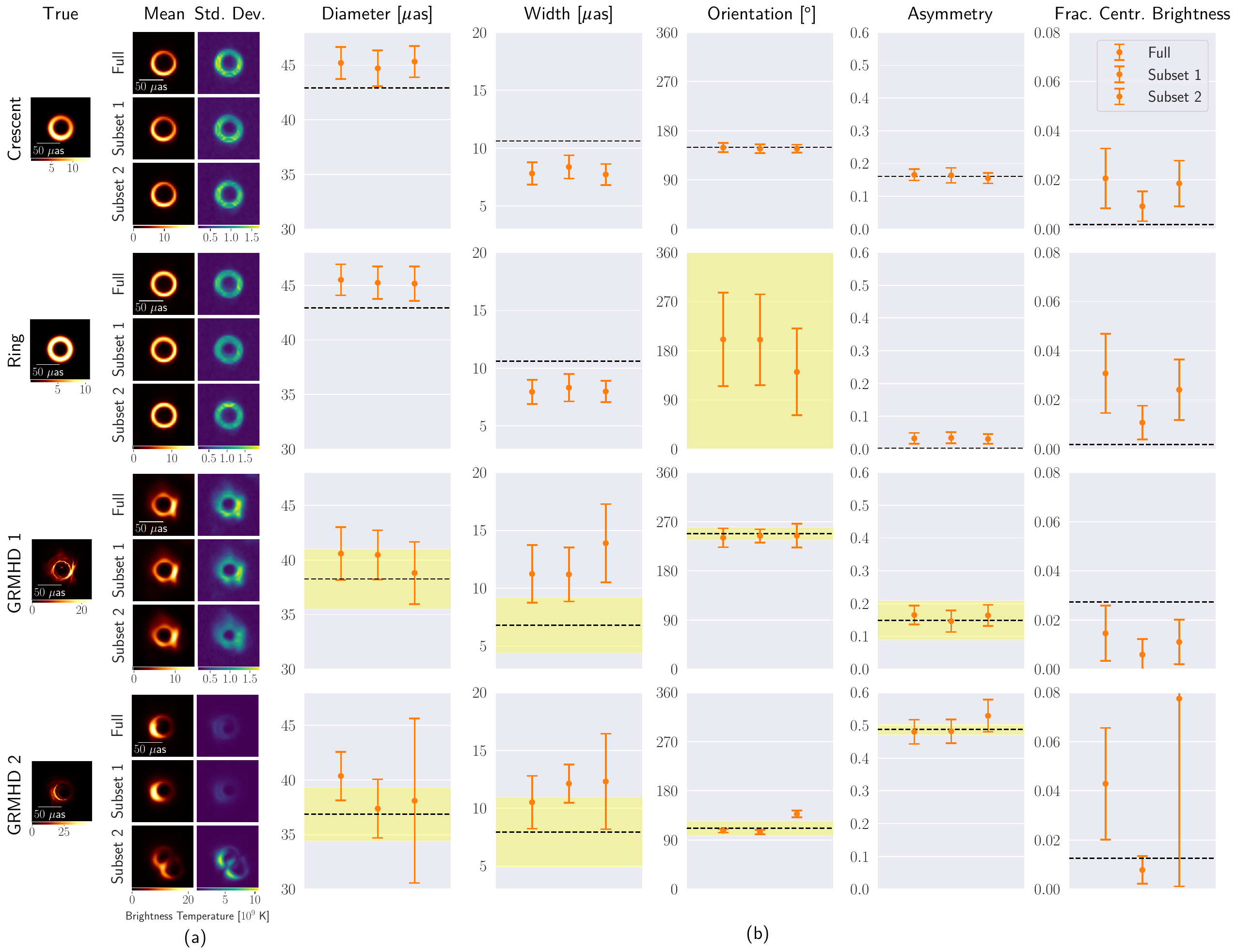}
    \caption{Model uncertainty of the GRMHD score-based prior. We trained a prior on the full GRMHD training dataset (``Full'') and two priors (``Subset 1'' and ``Subset 2'') on two disjoint subsets of the GRMHD training dataset. We obtained posterior samples under these three realizations of a GRMHD prior given synthetic observations of the four ring-like test images. Panel (a) shows the posterior means and pixel-wise standard deviations. The visual statistics of the three priors are similar. However, as the GRMHD 2 posterior under Subset 2 demonstrates, due to randomness in training, the resulting score-based prior may not always impose certain features so strongly. In this case, the Subset 2 prior results in some off-center rings when reconstructing the GRMHD 2 image. Panel (b) shows extracted ring features, comparing to the ground truth as shown in Figure \ref{fig:simdata_features}. Consistent with the results for the $64\times 64$ GRMHD prior in Figure \ref{fig:simdata_features}, these realizations of a $32\times 32$ GRMHD prior tend to slightly overestimate the diameter. However, among the different realizations of the prior, the extracted ring features are fairly consistent. The feature most sensitive to model uncertainty of the prior is fractional central brightness.}
    \label{fig:prior_uncertainty}
\end{figure}

In this appendix, we explore model uncertainty of the score-based prior and how it affects estimated ring features. We performed an experiment in which we divided the GRMHD training dataset into two subsets of $50$K images and trained a score-based prior on each subset, as well as a score-based prior on the full dataset. For this experiment, the GRMHD images were resized to $32\times 32$. Model uncertainty arises from the different training images and randomness during training, including randomness of data augmentations. We applied these three realizations of a GRMHD prior to the four ring-like test images (using a slower data annealing rate than the one used in the main results). Figure \ref{fig:prior_uncertainty} compares results of the three versions (``Full,'' ``Subset 1,'' and ``Subset 2'') of the GRMHD prior. The mean reconstructions are generally very similar, although it is possible for a version of the prior to impose the centering preference less strongly (see the Subset 2 posterior for the GRMHD 2 test image, which includes a mode that has an off-center ring). We also find that the extracted ring features are fairly robust to model uncertainty of the prior, as the error bars of the extracted features under the three different realizations of the prior usually overlap.

\section{Score-based Priors}
\label{app:score}
To obtain a score-based prior, we train a score-based diffusion model with parameters $\theta$ on samples from the target prior $p_\text{data}$. Once trained, the diffusion model can sample from a learned distribution $p_\theta\approx p_\text{data}$. In this appendix, we provide technical background on score-based diffusion models and how to evaluate $p_\theta$.

\subsection{Score-based Diffusion Models}
A score-based diffusion model learns to gradually denoise samples from the tractable distribution $\pi=\mathcal{N}(\mathbf{0},\mathbf{I})$ into samples from a clean image distribution $p_\text{data}$. Denoising is treated as a continuous-time process such that an image is a time-dependent variable $\mathbf{x}(t)$ for $t\in[0,T]$. A higher $t$ corresponds to higher independent Gaussian noise in $\mathbf{x}(t)$. The following stochastic differential equation (SDE) dictates how $\mathbf{x}(t)$ is ``diffused'' as $t$ goes from $0$ to $T$:
\begin{align}
\label{eq:forward_sde}
    \mathrm{d}\mathbf{x}=\mathbf{f}(\mathbf{x},t)+g(t)\mathrm{d}\mathbf{w},\quad t\in[0,T].
\end{align}
Here $\mathbf{w}\in\mathbb{R}^D$ is a standard Brownian motion, $\mathbf{f}(\cdot,t):\mathbb{R}^D\to\mathbb{R}^D$ is the drift coefficient that controls the deterministic evolution of $\mathbf{x}(t)$, and $g(t)\in\mathbb{R}$ is the diffusion coefficient that controls the rate of noise increase in $\mathbf{x}(t)$. (Note that $\mathbf{f}(\cdot,t)$ here is distinct from the forward model $\mathbf{f}(\cdot)$ in Section \ref{subsec:bayesian_imaging}.)
The stochastic trajectory $\{\mathbf{x}(t)\}_{t=0}^T$ gives rise to a time-dependent marginal probability distribution $p_t$. In addition, $\mathbf{f}(\cdot,t)$ and $g(t)$ are constructed such that if $p_0=p_\text{data}$, then $p_T\approx\pi$.

A diffusion model learns to \textit{reverse} the diffusion process. One can sample from the complicated distribution $p_\text{data}$ by first sampling from the tractable $\pi$ distribution and then using the diffusion model to transform those samples into clean images from $p_\text{data}$. The reverse of the diffusion process is given by the following reverse-time SDE:
\begin{align}
\label{eq:reverse_sde}
    \mathrm{d}\mathbf{x}=\left[\mathbf{f}(\mathbf{x},t)-g(t)^2\nabla_{\mathbf{x}}\log p_t(\mathbf{x})\right]\mathrm{d}t+g(t)\mathrm{d}\bar{\mathbf{w}},\quad t\in[0,T].
\end{align}
The gradient $\nabla_{\mathbf{x}}\log p_t(\mathbf{x})$ is known as the \textit{Stein score}~\citep{stein1972bound} of $\mathbf{x}$ under $p_t$, and it helps bring noisy images closer to the clean image distribution. It is also the component in Equation \eqref{eq:reverse_sde} that is learned by the diffusion model. A convolutional neural network with parameters $\theta$, known as the \textit{score model} $\mathbf{s}_\theta$, is trained to approximate the true score such that $\mathbf{s}_\theta(\mathbf{x},t)\approx\nabla_{\mathbf{x}}\log p_t(\mathbf{x})$. The score model appears in the following learned approximation of Equation \eqref{eq:reverse_sde}, which is used to sample from an approximation of $p_\text{data}$:
\begin{align}
\label{eq:learned_reverse_sde}
    \mathrm{d}\mathbf{x}=\left[\mathbf{f}(\mathbf{x},t)-g(t)^2\mathbf{s}_\theta(\mathbf{x},t)\right]\mathrm{d}t+g(t)\mathrm{d}\bar{\mathbf{w}},\quad t\in[0,T].
\end{align}
For $\mathbf{x}(T)\sim\pi$, we denote the time-dependent marginal distribution of $\mathbf{x}(t)$ according to Equation \eqref{eq:learned_reverse_sde} as $\hat{p}_t$. We denote the diffusion-model prior as $p_\theta := \hat{p}_0$.
For a well-trained score model, we have that $p_\theta\approx p_\text{data}$.

\subsection{Image Probabilities under a Score-based Diffusion Model}
To use $p_\theta$ as a prior in an inference algorithm that optimizes the posterior log density, we need access to the function $\log p_\theta(\cdot)$. Computing the probability of an image $\mathbf{x}$ under $p_\theta$ requires inverting $\mathbf{x}(0)=\mathbf{x}$ to $\mathbf{x}(T)$ (i.e., we need to find the $\mathbf{x}(T)$ that would result in $\mathbf{x}(0)$ through the reverse diffusion defined by Equation \eqref{eq:learned_reverse_sde}). However, although we can use Equation \eqref{eq:learned_reverse_sde} to sample from $p_\theta$, the presence of Brownian motion makes the sampling function not invertible. As a result, there is no tractable way to compute $\log p_\theta(\mathbf{x})$. Instead, we can appeal to an ordinary differential equation (ODE) for tractable log-probabilities or to the ELBO as an efficient proxy.

\subsubsection{Computing Probabilities with an ODE}
The following \textit{probability flow ODE} defines a generative image distribution $p_\theta^\text{ODE}$ theoretically equal to $p_\theta$:
\begin{align}
\label{eq:ode}
\frac{\mathrm{d}\mathbf{x}}{\mathrm{d}t}=\mathbf{f}(\mathbf{x},t)-\frac{1}{2}g(t)^2\mathbf{s}_\theta(\mathbf{x},t)=:\tilde{\mathbf{f}}_\theta(\mathbf{x},t).
\end{align}
Like the reverse-time SDE, this ODE can be used to sample $\mathbf{x}(0)\sim p_\theta^\text{ODE}$ by starting with $\mathbf{x}(T)\sim\pi$. Furthermore, an ODE can be solved in both directions of time, making the sampling function invertible. To compute $\log p_\theta^\text{ODE}(\mathbf{x})$ for an image $\mathbf{x}$, we map $\mathbf{x}(0)=\mathbf{x}$ to $\mathbf{x}(T)$ by solving the ODE in the forward time direction. The log probability is given by the log probability of $\mathbf{x}(T)$ under $\pi$ (which is tractable to evaluate) plus a normalization factor accounting for the change in log density throughout the ODE:
\begin{align}
\label{eq:ode_logprob}
    \log p_\theta^\text{ODE}(\mathbf{x}(0))=\log\pi(\mathbf{x}(T))+\int_0^T\nabla\cdot\tilde{\mathbf{f}}_\theta(\mathbf{x}(t),t)\mathrm{d}t,\quad\mathbf{x}(0)=\mathbf{x}.
\end{align}
Equation \eqref{eq:ode_logprob} is tractable to evaluate with an ODE solver, but it is computationally expensive. It can be prohibitively expensive when the image is large or when used in an iterative optimization algorithm.

\subsubsection{Surrogate Probabilities with an Evidence Lower Bound}
\citet{song2021maximum} derived the ELBO $b_\theta$ of a score-based diffusion model such that $b_\theta(\mathbf{x})\leq\log p_\theta(\mathbf{x})$ for any $\mathbf{x}$. The lower bound, which is similar to the denoising-based training objective of diffusion models, is given by
\begin{align}
\label{eq:dsm}
b_\theta&(\mathbf{x}):=\mathbb{E}_{p_{0T}(\mathbf{x}'\mid\mathbf{x})}\left[\log \pi (\mathbf{x}')\right] -\frac{1}{2}\int_0^T g(t)^2h(\mathbf{x},t)\mathrm{d}t,
\end{align}
where
% \small
\begin{align}
\label{eq:dsm_ht}
    h(\mathbf{x},t):= \mathbb{E}_{p_{0t}(\mathbf{x}'\mid\mathbf{x})}\left[\left\lVert\mathbf{s}_\theta(\mathbf{x}',t)-\nabla_{\mathbf{x}'}\log p_{0t}(\mathbf{x}'\mid\mathbf{x})\right\rVert_2^2 -\left\lVert\nabla_{\mathbf{x}'}\log p_{0t}(\mathbf{x}'\mid\mathbf{x})\right\rVert_2^2-\frac{2}{g(t)^2}\nabla_{\mathbf{x}'}\cdot\mathbf{f}(\mathbf{x}',t)\right].
\end{align}
% \normalsize
Here $p_{0t}(\mathbf{x}'\mid\mathbf{x})$ denotes the transition distribution of $\mathbf{x}(t)=\mathbf{x}'$ given $\mathbf{x}(0)=\mathbf{x}$ under the diffusion SDE (Equation \eqref{eq:forward_sde}). In denoising diffusion models, which use a linear drift coefficient $\mathbf{f}(\cdot,t)$, this transition distribution is Gaussian: $p_{0t}(\mathbf{x}'\mid\mathbf{x}) = \mathcal{N}(\mathbf{x}'; \alpha(t)\mathbf{x}, \beta(t)^2\mathbf{I})$. This in turn makes $\int_0^T g(t)^2h(\mathbf{x},t)\mathrm{d}t$ similar to a denoising score-matching objective. In fact, Equation \eqref{eq:dsm} is known as the ``denoising score-matching loss'' in the original theoretical work \citep{song2021maximum}.

To see the correspondence to denoising, note that for $p_{0t}(\mathbf{x}'\mid\mathbf{x}) = \mathcal{N}(\mathbf{x}'; \alpha(t)\mathbf{x}, \beta(t)^2\mathbf{I})$ we can sample $\mathbf{x}'$ as $\alpha_t\mathbf{x}+\beta_t\mathbf{z}$ for $\mathbf{z}\sim\mathcal{N}(\mathbf{0},\mathbf{I})$. Then the gradient $\nabla_{\mathbf{x}'}\log p_{0t}(\mathbf{x}'\mid\mathbf{x})$ in Equation \eqref{eq:dsm_ht} is given by
\begin{align}
    \nabla_{\mathbf{x}'}\log p_{0t}(\mathbf{x}'\mid\mathbf{x})=-\frac{1}{\beta_t^2}(\mathbf{x}'-\alpha_t\mathbf{x})=-\frac{\mathbf{z}}{\beta_t}.
\end{align}
The first term in Equation \eqref{eq:dsm_ht}, $\left\lVert\mathbf{s}_\theta(\mathbf{x}',t)-\nabla_{\mathbf{x}'}\log p_{0t}(\mathbf{x}'\mid\mathbf{x})\right\rVert_2^2$, thus can be thought-of as the denoising error of the score model. A lower error corresponds to a higher ELBO according to Equation \eqref{eq:dsm}. The remaining two terms in Equation \eqref{eq:dsm_ht} are normalizing factors independent of $\theta$. Loosely speaking, the ELBO indicates how well the diffusion model can denoise the given image: an image with high probability under the diffusion model is easy to denoise, whereas a low-probability image is difficult to denoise.
% REWORD: The term $\mathbb{E}_{p_{0T}(\mathbf{x}'\mid\mathbf{x})}\left[\log \pi (\mathbf{x}')\right]$ accounts for the probabilities of the noise images $\mathbf{x}(T)$ that could result from $\mathbf{x}$ being entirely diffused.

The quantity $b_\theta(\mathbf{x})$ is efficient to compute by adding Gaussian noise to $\mathbf{x}$ and seeing how well the score model estimates the noise. We Monte-Carlo approximate the expectation over $p_{0T}(\mathbf{x}'\mid\mathbf{x})$ with $N_\mathbf{z}$ noise samples, and we Monte-Carlo approximate the time integral in Equation \eqref{eq:dsm} with importance sampling of $N_t$ time samples. In practice, $N_\mathbf{z}=N_t=1$ is sufficient for our imaging algorithm.

\section{Interferometric Data Products}
\label{app:interferometry}
In this appendix, we provide background on the data products obtained with VLBI.
In VLBI, a network of radio telescopes collects spatial-frequency measurements of the sky's image. We denote the source image as $I(x,y)$, where $(x,y)$ are 2D spatial coordinates. Each pair of telescopes is called a \textit{baseline} and provides a Fourier measurement called a \textit{visibility}. The van Cittert-Zernike Theorem~\citep{van1934wahrscheinliche,zernike1938concept} states that the ideal visibility $v^*_{ij}$ measured by the baseline $\mathbf{b}_{ij}$ between telescopes $i$ and $j$ is a single $(u,v)$ measurement on the complex 2D Fourier plane~\citep{thompson2017interferometry}:
\begin{align}
\label{eq:ideal_vis}
    v^*_{ij}:=\tilde{I}(u,v)=\int\int I(x,y)e^{-2\pi \text{i}(xu +yv)}\mathrm{d}x \mathrm{d}y.
\end{align}
(Here i is used to denote the imaginary unit to avoid confusion with the telescope index $i$.) The coordinates $(u,v)$ (measured in wavelengths) are the projected baseline orthogonal to the line of sight.
An array of $N_s$ telescopes, or stations, has 
$\binom{N_s}{2}$ independent baselines, each providing a visibility at each point in time.

In practice, ideal visibilities are corrupted due to multiple factors: (1) baseline-dependent thermal noise, (2) station-dependent gain errors, and (3) station-dependent phase errors. Baseline-dependent \textbf{thermal noise} is modeled as a Gaussian random variable $\varepsilon_{ij}\sim\mathcal{N}(0,\sigma_{ij}^2)$, where $\sigma_{ij}$ is based on the system equivalent flux density (SEFD) of each telescope: $\sigma_{ij}\propto\sqrt{\text{SEFD}_i+\text{SEFD}_j}$. The \textbf{station-dependent gain error} $g_i$ arises from each telescope $i$ using its own time-dependent $2\times 2$ Jones matrix~{\citep{hamaker1996understanding}. The \textbf{station-dependent phase error} $\phi_i$ arises from atmostpheric turbulence that causes light to travel at different velocities toward each telescope~\citep{hinder1970observations,kolmogorov1991local,taylor1938spectrum}.
Other sources of corruption, including polarization leakage and bandpass errors, may introduce baseline-dependent errors, but they are slow-varying and assumed to be removable with a priori calibration~\citep{chael2018interferometric}. The measured visibility of baseline $\mathbf{b}_{ij}$ can be written as
\begin{align}
\label{eq:meas_vis}
    v_{ij} = g_ig_j e^{\text{i}(\phi_i-\phi_j)}v^*_{ij}+\varepsilon_{ij},
\end{align}
where all systematic errors (i.e., those besides thermal noise) are wrapped into station-dependent gain/phase errors.

\subsection{Closure Quantities}
Station-dependent errors are difficult to remove owing to the absence of corroborating information from other stations. Calibrating the measured visibilities calls for an iterative self-calibration process that introduces many a priori assumptions and becomes infeasible at high telescope frequencies~\citep{chael2018interferometric}. An alternative avenue is to use closure quantities that are unchanged by station-dependent errors.

\textbf{Closure phases} \citep{jennison1958phase} are robust to station-dependent phase errors. They arise from a data product known as the complex \textit{bispectrum}, which is formed by multiplying the three baselines within a triangle of telescopes $i,j,k$:
\begin{align}
    v_{ij}v_{jk}v_{ki}&=\left(g_ig_j e^{\text{i}(\phi_i-\phi_j)}v^*_{ij}+\varepsilon_{ij}\right)\left(g_jg_k e^{\text{i}(\phi_j-\phi_k)}v^*_{jk}+\varepsilon_{jk}\right)\left(g_kg_i e^{\text{i}(\phi_k-\phi_j)}v^*_{ki}+\varepsilon_{ki}\right) \\
    &=g_{ijk}^2 e^{\text{i}(\phi_i-\phi_j)}e^{\text{i}(\phi_j-\phi_k)}e^{\text{i}(\phi_k-\phi_i)}v^*_{ij}v^*_{jk}v^*_{ki}+\varepsilon_{ijk} \\
    &=g_{ijk}^2v^*_{ij}v^*_{jk}v^*_{ki}+\varepsilon_{ijk},
    \label{eq:bispectrum}
\end{align}
where $\varepsilon_{ijk}$ is the thermal noise in the measured bispectrum. Importantly, Equation \eqref{eq:bispectrum} does not include any phase errors. Thus the closure phase is given by the phase of the bispectrum and is robust to phase corruption. While the total number of triplets in the telescope array is $\binom{N_s}{3}$, the total number of linearly independent closure phases is $N_{\text{cp}}=\binom{N_s-1}{2}$. To understand this number, see that the set of independent closure phases can be formed by selecting one station as a reference and then creating the set of all triangles that contain that station. The resulting closure phases are independent in that no one closure phase can be formed as a linear combination of other closure phases. 

\textbf{Closure amplitudes} \citep{twiss1960brightness} address the issue of station-dependent gain errors. A closure amplitude arises from a combination of four telescopes $i,j,k,\ell$:
\begin{align}
    \frac{v_{ij}v_{kl}}{v_{ik}v_{jl}}&=\frac{\left(g_ig_je^{\text{i}(\phi_i-\phi_j)}v^*_{ij}+\varepsilon_{ij}\right)\left(g_kg_\ell e^{\text{i}(\phi_k-\phi_\ell)}v^*_{k\ell}+\varepsilon_{k\ell}\right)}{\left(g_ig_ke^{\text{i}(\phi_i-\phi_k)}v^*_{ik}+\varepsilon_{ik}\right)\left(g_jg_\ell e^{\text{i}(\phi_j-\phi_\ell)}v^*_{j\ell}+\varepsilon_{j\ell}\right)} \\
    &= \frac{g_ig_jg_kg_\ell e^{\text{i}(\phi_i-\phi_j)}e^{\text{i}(\phi_k-\phi_\ell)}v^*_{ij}v^*_{k\ell}}{g_ig_jg_kg_\ell e^{\text{i}(\phi_i-\phi_k)}e^{\text{i}(\phi_j-\phi_\ell)}v^*_{ik}v^*_{j\ell}}+\varepsilon_{ijk\ell} \\
    &= e^{2\text{i}(\phi_k-\phi_j)}\frac{v^*_{ij}v^*_{k\ell}}{v^*_{ik}v^*_{j\ell}}+\varepsilon_{ijk\ell}, \label{eq:logca_spectrum}
\end{align}
where $\varepsilon_{ijk\ell}$ is the thermal noise in the closure amplitude. Equation \eqref{eq:logca_spectrum} does not depend on station-dependent gain errors, so the amplitude $\left\lvert \frac{v_{ij}v_{kl}}{v_{ik}v_{jl}} \right\rvert$ is taken as the closure amplitude. In this work, we use the log of the closure amplitude. The number of linearly independent log closure amplitudes is $N_\text{logca}=\binom{N_s(N_s-3)}{2}$.

% \subsection{EHT M87 observations}
% The EHT M87 observations were obtained on April 5, 6, 10, and 11 of 2017 with an array of seven stations at five different locations around the globe. The telescopes each measured two 2 GHz bands centered on 227.1 and 229.1 GHz. The observations were correlated, corrected, and reduced to form the final network-calibrated data sets~\citep{m87paperiii}.

% Starting with the published network-calibrated data sets, we follow the same data processing pipeline as the original M87 imaging work~\citep{m87paperiv}. In particular, we utilize the Stokes $I$ visibilities, combine measured visibilities from the two frequency bands, and coherently average the visibilities per 10-second interval. We then rescale short baselines to excise extended flux beyond the assumed compact flux of 0.6 Jy. 

\section{Ring Feature Extraction}
\label{app:feature_extraction}
We used the \texttt{REx} feature extraction algorithm \citep{chael2019simulating} to compute the characteristic features in Section \ref{sec:features}. The diameter and fractional central brightness follow the same formulae as in \citet{m87paperiv}. Except for the fractional central brightness, which is not included in \texttt{REx}, all features were computed exactly according to the latest implementation of \texttt{REx}\footnote{\url{https://github.com/achael/eht-imaging/blob/main/ehtim/features/rex.py}} (as of 2023 October). The \texttt{REx} implementation corresponds to slightly different equations for computing features than those given in \citet{m87paperiv}. In this appendix, we will note any differences from the equations used in \citet{m87paperiv}.

\texttt{REx} first pre-processes the image by blurring it with a $2$ $\mu$as FWHM Gaussian and re-gridding it to $160\times 160$ pixels. It identifies the ring center based on the image thresholded to $5\%$ of the peak brightness, and then it computes characteristic features. The characteristic features, which we define in the following paragraphs, are all computed based on radial--angular profiles $I(r,\theta)$ of the centered image, where $I(r,\theta)$ is the brightness at radius $r$ and azimuthal angle $\theta$ from the measured center. The profiles are interpolated over the domains $r\in[0,50]$ $\mu$as and $\theta\in[0,2\pi]$ radians.

The diameter $d$ is measured as twice the mean radial distance of the peak brightness:
\begin{align}
    d = 2 \left\langle r_\text{pk}(\theta) \right\rangle_{\theta\in[0,2\pi]} := 2 \left\langle \arg\max_r I(r,\theta) \right\rangle_{\theta\in[0,2\pi]},
\end{align}
where $\langle\cdot\rangle_{\theta\in[0,2\pi]}$ denotes the mean over the domain $\theta\in[0,2\pi]$. The uncertainty of the diameter is given as the corresponding standard deviation. This equation for $d$ exactly agrees with Equation (18) in \citet{m87paperiv}.

The width $w$ is measured as the mean FWHM of radial slices:
\begin{align}
    w=\left\langle \text{FWHM}\left(I(r,\theta)\right) \right\rangle_{\theta\in[0,2\pi]},
\end{align}
where $\text{FWHM}(\cdot)$ evaluates the FWHM of a 1D radial profile. The uncertainty is computed as the corresponding standard deviation. Note that this is slightly different from the expression for $w$ given in Equation (20) in \citet{m87paperiv}, which first subtracts the mean flux outside the ring before estimating the width.

To measure the orientation angle, \texttt{REx} first estimates the FWHM of the mean radial profile with zero-mean outside flux, defined as
$$\text{FWHM}\left(\bar{I}(r)-I_\text{floor}\right),$$ where $\bar{I}(r):=\langle I(r, \theta) \rangle_{\theta\in[0,2\pi]}$ and $I_\text{floor}:=\left\langle I(r=50,\theta)\right\rangle_{\theta\in[0,2\pi]}$. Let $r_\text{left}$ and $r_\text{right}$ denote the minimum and maximum radii, respectively, of this FWHM. The orientation angle is computed by finding the phase of the first angular mode of each normalized angular profile $I(r,\theta)$ at fixed $r$ throughout the FWHM and then taking the circular mean:
\begin{align}
    \eta = \left\langle \angle \left[\frac{\int_0^{2\pi}I(r,\theta)e^{i\theta}\mathrm{d}\theta}{\int_0^{2\pi}I(r,\theta)\mathrm{d}\theta}\right]\right\rangle_{r\in[r_\text{left},r_\text{right}]}.
\end{align}
The uncertainty of $\eta$ is the corresponding circular standard deviation. The only difference between this equation and Equation (21) in \citet{m87paperiv} comes from $r_\text{left}$ and $r_\text{right}$. In \citet{m87paperiv}, $r_\text{in}=(d-w)/2$ and $r_\text{out}=(d+w)/2$ are used instead.

The azimuthal asymmetry $A$ is measured as the mean normalized amplitude of the same first angular modes:
\begin{align}
    A = \left\langle \left\lvert\frac{\int_0^{2\pi}I(r,\theta)e^{i\theta}\mathrm{d}\theta}{\int_0^{2\pi}I(r,\theta)\mathrm{d}\theta}\right\rvert\right\rangle_{r\in[r_\text{left},r_\text{right}]},
\end{align}
and the uncertainty of $A$ is the corresponding standard deviation. Once again, the only difference from Equation (22) in \citet{m87paperiv} is that $r_\text{left}$ and $r_\text{right}$ are used instead of $r_\text{in}$ and $r_\text{out}$.

We define the fractional central brightness $f_\text{C}$ (which is not included in \texttt{REx}) as the ratio of an ``interior'' mean flux to the mean flux along the ring:
\begin{align}
    f_\text{C}=\frac{\left\langle I(r,\theta)\right\rangle_{r\in[0,5],\theta\in[0,2\pi]}}{\left\langle I(d/2,\theta) \right\rangle_{\theta\in[0,2\pi]}}.
\end{align}
Here the inside is defined as the inner disk of radius $5$ $\mu$as, and the outside is defined as the region with radius larger than the measured radius $d/2$. There is no uncertainty quantification for $f_\text{C}$. This definition of $f_\text{C}$ exactly agrees with Equation (23) in \citet{m87paperiv}.

\section{Implementation Details}
\label{app:implementation}
In this appendix, we provide implementation details about the approach used in our experiments.

\subsection{Optimization of the RealNVP Variational Posterior}
For the variational distribution, we used a RealNVP normalizing-flow network with $32$ affine-coupling layers with a width of $D/8$ neurons in the first layer, where $D$ is the number of pixels in the images (i.e., $32\times 32$ or $64\times 64$ in this work). Stochastic gradient-based optimization was done with batches of $64$ images and the Adam optimizer with a learning rate of $10^{-5}$ and gradient clip of $1$. For each posterior, we ran optimization for $100$K iterations.
% Images generated by a RealNVP might have some slightly negative pixel values, so at inference time, we zero-clip negative values. Zero-clipped samples from the optimized RealNVP are taken as approximate posterior samples. 

\subsection{Score-based Prior Settings}
For the score-based diffusion model, we used a score model with the NCSN++ architecture with $64$ filters in the first layer, trained according to the Variance-Preserving (VP) SDE~\citep{song2021scorebased}. The CIFAR-10, GRMHD, RIAF, and CelebA score-based priors were trained for $100$K, $100$K, $20$K, and $100$K iterations, respectively.
The ELBO $b_\theta(\mathbf{x})$ (Equation \eqref{eq:dsm}) was Monte-Carlo-approximated with $N_t=1$ time sample and $N_\mathbf{z}=1$ noise sample.

\bibliography{main}{}
\bibliographystyle{aasjournal}


\listofchanges

\end{document}